{}%disable hyper at arxiv
{}%use pdflatex at arxiv
 
\documentclass[11pt,a4paper]{article}
\newif\ifpublic\publictrue
\newif\iffancy\fancytrue

\usepackage[a4paper,text={160mm,247mm},centering]{geometry}
\usepackage{setspace}
\usepackage{amsmath,amssymb, amsfonts, geometry, shuffle}
\usepackage[pdfencoding=auto,bookmarks=true,hyperfigures=true]{hyperref}
\usepackage{graphicx}% uncomment for latex-use
\usepackage{float}
\usepackage[nosort]{cite}
\usepackage{lmodern}
\usepackage{amsbsy}
\usepackage[utf8]{inputenc}
\usepackage[font=small,labelfont=bf]{caption}

\usepackage[usenames,dvipsnames]{xcolor}
\definecolor{dgreen}{rgb}{0,0.70,0.30}
\definecolor{gold}{rgb}{0.85,.66,0}
\definecolor{purple}{rgb}{1.0,0.3,0.6}
\usepackage{tikz}
\usetikzlibrary{calc} 
\usetikzlibrary{patterns} 
\usetikzlibrary{decorations.pathreplacing} 
\usetikzlibrary{decorations.markings} 
\usetikzlibrary{decorations.pathmorphing} 
\usetikzlibrary{positioning}

\iffancy

\def\showkeysrefformat#1{{\normalfont\tiny\ttfamily#1}}
\makeatletter
\def\SK@@ref#1>#2\SK@{%
 {\@inlabelfalse\leavevmode\vbox to\z@{%
 \vss\SK@refcolor\rlap{\vrule\raise .75em%
  \hbox{\showkeysrefformat{#2}}}}}}
\makeatother
\fi

\numberwithin{equation}{section}

\providecommand{\href}[2]{#2}

\makeatletter
\def\mr@ignsp#1 {\ifx\:#1\@empty\else #1\expandafter\mr@ignsp\fi}%
\newcommand{\multiref}[1]{\begingroup%\let\protect\string%
\xdef\mr@no@sparg{\expandafter\mr@ignsp#1 \: }%
\def\mr@comma{}%
\@for\mr@refs:=\mr@no@sparg\do{\mr@comma\def\mr@comma{,}\ref{\mr@refs}}%
\endgroup}
\renewcommand{\eqref}[1]{(\multiref{#1})}
\makeatother

\makeatletter
\newcommand{\namedref}[2]{\hyperref[#2]{#1~\ref*{#2}}}
\newcommand{\secref}{\@ifstar{\namedref{Section}}{\namedref{section}}}
\newcommand{\subsecref}{\@ifstar{\namedref{Subsection}}{\namedref{subsection}}}
\newcommand{\appref}{\@ifstar{\namedref{Appendix}}{\namedref{appendix}}}
\newcommand{\tabref}{\@ifstar{\namedref{Table}}{\namedref{table}}}
\newcommand{\figref}{\@ifstar{\namedref{Figure}}{\namedref{figure}}}
\makeatother

\newcommand{\eqn}[1]{eq.~\eqref{#1}}

\newcommand{\eqns}[2]{eqs.~\eqref{#1} and~\eqref{#2}}

\providecommand{\hypersetup}[1]{}
\providecommand{\texorpdfstring}[2]{#1}

\hypersetup{plainpages=false}
\hypersetup{pdfpagemode=UseNone}
\hypersetup{bookmarksnumbered=true}
\hypersetup{pdfstartview=FitH}
\hypersetup{colorlinks=false}
\hypersetup{citebordercolor={.5 1 .5}}
\hypersetup{urlbordercolor={.5 1 1}}
\hypersetup{linkbordercolor={1 .7 .7}}
\makeatletter
\let\@keywords\@empty
\let\@subject\@empty
\providecommand{\keywords}[1]{\gdef\@keywords{#1}}
\providecommand{\subject}[1]{\gdef\@subject{#1}}
\def\thetitle{\@title}
\def\theauthor{\@author}
\def\thesubject{\@subject}
\def\thedate{\@date}
\def\thekeywords{\@keywords}
\makeatother
\AtBeginDocument{
\hypersetup{pdftitle={\thetitle}}%
\hypersetup{pdfauthor={\theauthor}}%
\hypersetup{pdfsubject={\thesubject}}%
\hypersetup{pdfkeywords={\thekeywords}}%
}

\RequirePackage{verbatim}

\newif\ifnote 
\notetrue

\allowdisplaybreaks

\newcommand{\ba}  {\begin{array}}
\newcommand{\ea}  {\end{array}}
\newcommand{\bdm} {\begin{displaymath}}
\newcommand{\edm} {\end{displaymath}}

\newcommand{\bea} {\begin{equation}\ba{lcl}}
\newcommand{\eea} {\ea\end{equation}}

\newcommand{\bc}  {\begin{center}}
\newcommand{\ec}  {\end{center}}

\newcommand{\p} {\partial}
\newcommand{\ad} {\mathrm{ad}}

\def\beq{\begin{equation}}
\def\eeq{\end{equation}}

\def\Re{{\rm Re\,}}
\def\Im{{\rm Im\,}}

\newcommand{\vecb}{\left(\begin{array}{c}}
\newcommand{\vece}{\end{array}\right)}
\newcommand{\ccb}{\left(\begin{array}{cc}}
\newcommand{\cce}{\end{array}\right)}
\newcommand{\cccb}{\left(\begin{array}{ccc}}
\newcommand{\ccce}{\end{array}\right)}
\newcommand{\ccccb}{\left(\begin{array}{cccc}}
\newcommand{\cccce}{\end{array}\right)}
\newcommand{\cccccb}{\left(\begin{array}{ccccc}}
\newcommand{\ccccce}{\end{array}\right)}
\newcommand{\pa}{\partial}
\newcommand{\pd}{\partial}

\def\a{\alpha}

\newcommand{\de}{\delta}

\newcommand{\z}{\zeta}

\newcommand{\s}{\sigma}

\newcommand{\te}{\textrm}
\newcommand{\co}{\ , \ \ \ \ \ \ }
\newcommand{\dd}{\mathrm{d}}

\newcommand{\half}{\tfrac{1}{2}}

\newcommand{\ap}{\alpha'}

\newcommand{\ZR}{\mathbb R}

\newcommand{\ZC}{\mathbb C}

\newcommand{\nnl}{\nonumber\\}

\DeclareMathOperator{\GL}{\Gamma}
\DeclareMathOperator{\ELi}{ELi}
\newcommand{\nol}{N}

\newcommand{\GLargz}[2]{\GL\left(\begin{smallmatrix}#1\\#2\end{smallmatrix};z\right)}
\newcommand{\GLarg}[3]{\GL\left(\begin{smallmatrix}#1\\#2\end{smallmatrix};#3\right)}
\newcommand{\IIECl}{elliptic iterated integrals}
\newcommand{\IIEC}{eII}

\def\f#1{f^{(#1)}}
\def\cE{{\cal E}}

\usepackage{amsthm} % fixes a compilation error
\theoremstyle{plain}

\title{\textbf{Elliptic multiple zeta values and\\one-loop superstring amplitudes}}
\author{Johannes Broedel$^{\te{a}}$, Carlos R.~Mafra$^{\te{b}}$, Nils Matthes$^{\te{c}}$, 
Oliver Schlotterer$^{\te{d}}$}
\date{\today}

\begin{document}
\pdfbookmark[1]{Title Page}{title} \thispagestyle{empty}
\begin{flushright}
  \verb!AEI-2014-066!\\
  \verb!DAMTP-2014-95!
\end{flushright}
\vspace*{1.8cm}
\begin{center}%
  \begingroup\LARGE\bfseries\thetitle\par\endgroup
  \vspace{1.4cm}

\begingroup\large\theauthor\par\endgroup
\vspace{8mm}%

\begingroup\itshape
$^{\te{a}}$Institut f\"ur Theoretische Physik,\\
Eidgen\"ossische Technische Hochschule Z\"urich,\\
Wolfgang-Pauli-Strasse 27, 8093 Z\"urich, Switzerland
\par\endgroup
\vspace{4mm}
\begingroup\itshape
$^{\te{b}}$Department of Applied Mathematics and Theoretical Physics,\\
University of Cambridge, Wilberforce Road, \\
Cambridge CB3 0WA, United Kingdom
\par\endgroup
\vspace{4mm}
\begingroup\itshape
$^{\te{c}}$Fachbereich Mathematik, Universit\"at Hamburg,\\
Bundesstra{\ss}e 55, 20146 Hamburg, Germany
\par\endgroup
\vspace{4mm}
\begingroup\itshape
$^{\te{d}}$Max-Planck-Institut f\"ur Gravitationsphysik,\\
Albert-Einstein-Institut, \\
Am M\"uhlenberg 1, 14476 Potsdam, Germany
\par\endgroup

\vspace{0.8cm}

\begingroup\ttfamily
jbroedel@ethz.ch, c.r.mafra@damtp.cam.ac.uk,

nils.matthes@uni-hamburg.de, olivers@aei.mpg.de 
\par\endgroup

\vspace{1.2cm}

\bigskip
%\vspace{0.7cm}

\textbf{Abstract}\vspace{5mm}

\begin{minipage}{13.4cm}
We investigate iterated integrals on an elliptic curve, which are a natural
genus-one generalization of multiple polylogarithms. These iterated integrals
coincide with the multiple elliptic polylogarithms introduced by Brown and
Levin when constrained to the real line.  At unit argument they reduce to an
elliptic analogue of multiple zeta values, whose network of relations we start
to explore.  A simple and natural application of this framework are one-loop
scattering amplitudes in open superstring theory. In particular, elliptic
multiple zeta values are a suitable language to express their low energy limit.
Similar to the techniques available at tree-level, our formalism allows to
completely automatize the calculation.
\end{minipage}

%\vspace*{4cm}

\end{center}

\newpage

%\setcounter{tocdepth}{2}
%\tableofcontents

\section{Introduction}

In recent years, we have witnessed numerous fruitful interactions between
number theory and particle physics. A particularly rich domain of intersection
are iterated integrals, which prominently appear in scattering amplitudes in
field theories and string theories. For a large class of Feynman and worldsheet
integrals, multiple polylogarithms were recognized as a suitable language to
cast results into a manageable form, see
e.g.~refs.~\cite{Goncharov:2010jf,Dixon:2012yy,Broedel:2013tta,DelDuca:2013lma}.
In a variety of cases, the polylogarithms' Hopf algebra structure
\cite{Goncharov:2001iea, Goncharov:2005sla, Brown:2011ik, Duhr:2012fh} paved
the way towards efficient manipulations and the recognition of the simplicity
hidden in the resulting scattering amplitudes.

However, a growing list of iterated integrals from various field and string
theories implies that multiple polylogarithms do not mark the end of the rope
in terms of transcendental functions appearing in scattering amplitudes.  For
example, multiple polylogarithms fail to capture the two-loop sunset integral
with non-zero masses \cite{Adams:2013nia, Bloch:2013tra, Adams:2014vja}, an
eight-loop graph in $\phi^4$ theory \cite{Brown:kthree, Brown:2013wn} as well
as the ten-point two-loop N${}^3$MHV amplitude in ${\cal N}=4$
super-Yang--Mills (sYM) theory \cite{CaronHuot:2012ab}.  The sunset integral
and its generalization have recently been expressed in terms of elliptic di-
and trilogarithms \cite{Bloch:2013tra, Adams:2014vja, Bloch:2014qca}, whose
connection to the language suggested below remains to be worked out.
Considering in addition their appearance in one-loop open-string amplitudes,
the situation calls for a systematic study and classification of the entire
family of \IIECl\footnote{The elliptic iterated integrals discussed in this
work shall not be confused with elliptic integrals determining the arc length
of an ellipse.}.

In the present article, we propose a framework for \IIECl{} (or \IIEC s for
short) and the associated periods, elliptic multiple zeta values (eMZVs). The
framework aims at expressing scattering amplitudes in a variety of theories,
and we here apply the techniques to one-loop amplitudes in open string theory
as a first example. The language employed in the present article is primarily
inspired by refs.~\cite{BrownLev, Enriquez:emzv}, while
refs.~\cite{BeilinsonLevin,Bloch:2000,Levin,Wildeshaus,Zagier} contain further
information on the mathematical background.

As opposed to multiple polylogarithms, which can be defined using just one type
of differential form, \IIECl{} require an infinite tower thereof
\cite{BrownLev}. These differential forms are based on a certain
non-holomorphic extension of a classical Eisenstein-Kronecker series
\cite{Weil:1976,BrownLev}, and we show how they can be used to naturally
characterize and label \IIECl{} as well as eMZVs. We investigate their
relations, which results in constructive algorithms to perform amplitude
computations.

In the same way as multiple zeta values (MZVs) arise from multiple
polylogarithms at unit argument, the evaluation of iterated integrals along
a certain path of an elliptic curve leads to structurally interesting periods,
the eMZVs \cite{Enriquez:emzv} mentioned above.  These are certain analogues of
the standard MZVs, which are related to elliptic associators
\cite{Enriquez:ellass} in the same way as MZVs are related to
the Drinfeld associator \cite{Drinfeld:1989st, Drinfeld2, Le}. However, the
precise connection is beyond the scope of the current article. Given their
ubiquitous appearance in the subsequent string amplitude computation, we will
investigate eMZVs and discuss some of their properties as well as their
$\mathbb Q$-linear relations.

The description of string scattering amplitudes via punctured Riemann surfaces
at various genera directly leads to iterated integrals at the corresponding
loop order.  In particular, the disk integrals in open-string\footnote{In
comparison to open-string amplitudes at tree-level, MZVs occurring in
closed-string tree amplitudes \cite{Stieberger:2009rr, Schlotterer:2012ny} are
constrained by the single-valued projection, see \cite{Schnetz:2013hqa,
Brown:2013gia} for mathematics and \cite{Stieberger:2013wea,
Stieberger:2014hba} for physics literature.} tree-level amplitudes closely
resemble multiple polylogarithms.  Initially addressed via
hypergeometric functions in refs.~\cite{Oprisa:2005wu, Stieberger:2006te}, the
$\ap$-expansion of disk amplitudes finally proved to be a rich laboratory for
MZVs. Their pattern of appearance has been understood in terms of mathematical
structures such as motivic MZVs \cite{Brown:2011ik,Schlotterer:2012ny} and the
Drinfeld associator \cite{Terasoma, Drummond:2013vz, Broedel:2013aza}.
Explicit expressions with any number of open-string states can be determined
using polylogarithm manipulations \cite{Broedel:2013tta} or a matrix
representation of the associator \cite{Broedel:2013aza}. A variety of examples
are available for download at the website \cite{MZVWebsite}.

The calculation of one-loop open-string amplitudes involves worldsheet
integrals of cylinder and M\"obius-strip topology \cite{Green:1987mn}. In the
current article, we focus on iterated integrals over a single cylinder boundary
and leave the other topologies for later.  Recognizing the cylinder as a
genus-one surface with boundaries, it is not surprising that the
$\ap$-expansion of one-loop open-string amplitudes is a natural, simple and
representative framework for the application of \IIEC s and eMZVs. We will
explicitly perform calculations at four and five points for low orders in $\ap$
in order to demonstrate their usefulness. Higher multiplicities and orders in
$\ap$ are argued to yield eMZVs and Eisenstein series on general grounds. In
summary, one-loop string amplitudes turn out to be an ideal testing ground for
the study of eMZVs, in particular because they appear in a more digestible
context as compared to their instances in field theory.

This article is organized as follows: In \secref{sec:polylog}, we start by
reviewing multiple polylogarithms and show, how their structure suggests a
generalization to genus one. The appropriate differential forms and doubly-periodic
functions are discussed and put into a larger mathematical context in
\secref{sec:math}.  \secref*{sec:fourpoint} is devoted to the application of
\IIEC s and eMZVs to the four-point one-loop amplitude of the open string,
while \secref{sec:multipt} contains a discussion of its multi-particle
generalization.

%%%%%%%%%%%%%%%%%%%%%%%%%%%%%%%%%%%%%%%%%%%%%%%%%%%%%%%%%%%%%%%%%%%%%%%%%%%%%%%%

\section{Iterated integrals on an elliptic curve} 
\label{sec:polylog}

After recalling the definition of multiple polylogarithms as well as several
conventions, we will introduce \IIECl{} (\IIEC s)
as their genus-one analogues.  While we will limit ourselves to basic
definitions and calculational tools in the current section, a thorough
introduction to the mathematical background of doubly-periodic functions will be
provided in \secref{sec:math}. 

\subsection{Multiple polylogarithms}
\label{ssec:MP}
Multiple polylogarithms are defined by\footnote{The conventions for multiple
  polylogarithms used in this paper agree with those in
  refs.~\cite{Goncharov:2001iea, Schlotterer:2012ny, Duhr:2011zq}. Other aspects of multiple polylogarithms are discussed for example in references \cite{Ablinger:2013cf, Ablinger:2013jta}.} 
\begin{equation}
  G(a_1,a_2,\ldots,a_n;z) \equiv \int_0^z\frac{\dd t}{t-a_1}G(a_2,\ldots,a_n;t)
  \label{eqn:defG}
\end{equation}
where $G(;z)\equiv1$ apart from $G(\vec{a};0)=G(;0)=0$. Below, we will refer to
$\vec{a}=(a_1,\ldots,a_n)$ as the \textit{label} and call $z$ the
\textit{argument} of the polylogarithm $G$.  Powers of ordinary logarithms can
be conveniently represented in terms of multiple polylogarithms via
\begin{align}
  G(\underbrace{0,0,\ldots,0}_n;z)&=\frac{1}{n!}\ln^nz,\qquad\qquad
  G(\underbrace{1,1\ldots,1}_n;z) =  \frac{1}{n!}\ln^n(1-z)\quad\text{and}\nnl
  G(\underbrace{a,a,\ldots,a}_n;z)&=\frac{1}{n!}\ln^n\left(1-\frac{z}{a}\right)\,.
  \label{eqn:lntoHG}
\end{align}
In addition, multiple polylogarithms satisfy the scaling property
\begin{equation}
  G(ka_1,ka_2,\ldots,ka_n;kz)=G(a_1,a_2,\ldots,a_n;z)\co k \neq 0\ , \quad a_n \neq 0 \ , \quad z \neq 0 \,,
  \label{scaling}
\end{equation}
whose interplay with a general shuffle regularization will be
  discussed below \eqn{shufflereg2}. Another property is referred to as the
  H\"older convolution \cite{Borwein:1999js}: for $a_1\neq 1$ and $a_n\neq 0$ one
  finds
\begin{equation}
  G(a_1,\ldots,a_n;1)=\sum\limits_{k=0}^n(-1)^k\,G\bigg(1-a_k,\ldots,1-a_1;1-\frac{1}{p}\bigg)\,G\bigg(a_{k+1},\ldots,a_n;\frac{1}{p}\bigg)
  \label{eqn:Hoelder}
\end{equation}
for all $p\in\ZC\setminus\{0\}$. Multiple polylogarithms constitute a graded commutative
algebra with the shuffle product \cite{Goncharov:2001iea, Goncharov:2005sla,
Brown:2011ik, Duhr:2012fh}
\begin{align}
  G(a_1,\ldots,a_r;z)G(a_{r+1},\ldots,a_{r+s};z)&=\sum_{\s\in \Sigma(r,s)}G(a_{\s(1)},
\ldots,a_{\s(r+s)};z)
  \label{eqn:shuffle} \\
  &\equiv G\big( (a_1,\ldots,a_r) \shuffle (a_{r+1},\ldots,a_{r+s});z \big) \ ,
   \notag
\end{align}
where the shuffle $\Sigma(r,s)$ is the subset of the permutation group
$S_{r+s}$ acting on $\{a_1,\ldots,a_{r+s}\}$ which leaves the order of the
elements of the individual tuples $\{a_1, \ldots,a_r\}$ and
$\{a_{r+1},\ldots,a_{r+s}\}$ unchanged. The unit element for shuffling is
$G(;z)$=1.

MZVs are special cases of multiple polylogarithms with labels $a_i\in\{0,1\}$
evaluated at argument $z=1$:
\begin{equation}
  \zeta_{n_1,\ldots,n_r} = (-1)^r G(\underbrace{0,0,\ldots,0,1}_{n_r},\ldots,
\underbrace{0,0,\ldots,0,1}_{n_1};1)\,,
  \label{eqn:defZeta}
\end{equation}
where the numbers below the underbraces denote the number of
entries\footnote{Our convention for MZVs agrees with
  refs.~\cite{Goncharov:2001iea, Schlotterer:2012ny, BrownTate}.}. 

From the definition \eqref{eqn:defG} it is obvious that multiple polylogarithms
diverge when either $a_1=z$ or $a_n=0$. As discussed in
refs.~\cite{Goncharov:2001iea,Goncharov:2005sla}, the general idea for
regularizing the integrals is to slightly move the endpoints of the integration
by a small parameter and to afterwards expand in this parameter. The
regularized value of the polylogarithm is defined to be the piece independent
of the regularization parameter, which can be extracted using shuffle
relations. For the case where $a_1=z$ the regularized value can be obtained via 
\begin{align}
 G(z,a_2,\ldots,a_n;z)&= G(z;z)\,G(a_2,\ldots,a_n;z)-G(a_2,z,a_3,\ldots,a_n;z)\nnl
 &\qquad-G(a_2,a_3,z,a_4,\ldots,a_n;z)-\ldots-G(a_2,\ldots,a_n,z;z)
  \label{shufflereg1}
\end{align}
where one defines 
\begin{equation}
 G(z,\ldots,z;z)=0  \ .
  \label{shufflereg1a}
\end{equation}
The situation, where $a_n=0$ can be
dealt with accordingly 
\begin{align}
  G(a_1,a_2,\ldots,&a_{n-1},0;z)= G(a_1,a_2,\ldots,a_{n-1};z)\,G(0;z)-G(a_1,a_2,\ldots,
0,a_{n-1};z)\nnl
  &- G(a_1,a_2,\ldots,0,a_{n-2},a_{n-1};z)-\ldots-G(0,a_2,\ldots,a_{n-1};z)\,,
  \label{shufflereg2}
\end{align}
where now, however, $G(0;z) = \ln (z) \neq 0$. Although the above rewriting
keeps the pure logarithms explicit, it will nevertheless prove convenient in
order to bypass subtleties of the identity \eqn{eqn:GIdentity} below.  Multiple
polylogarithms are understood to be shuffle-regularized in a way compatible
with \eqn{scaling}. 

Regularization of multiple polylogarithms can be straightforwardly translated
to MZVs. All MZVs $\zeta_{n_1,\ldots,n_r}$ with $n_r=1$ are defined by their
shuffled version \eqn{shufflereg1}. Employing \eqn{scaling}, one finds
$G(1,\ldots,1;1)=0$ from \eqn{shufflereg1a} immediately. 

\subsubsection{Removing the argument \texorpdfstring{$\boldsymbol{z}$}{z} from the label} 

Starting from an arbitrary iterated integral, the corresponding
polylogarithm can not always be determined straightforwardly: whenever the
argument appears in the label $\vec{a}$, an integration using \eqn{eqn:defG} is
impossible.  Solving this problem requires a rewriting of the multiple
polylogarithm
\begin{equation}
  G(\{0,a_1,a_2,\ldots,a_n,z\};z)
  \label{eqn:zArg}
\end{equation}
in terms of polylogarithms whose labels are free of the argument. In the above
equation $\{a,b,\ldots\}$ refers to a word built from the letters
$a,b,\ldots\,\,\,$.  Polylogarithms of the special form $G(\vec{a},z)$ with
$a_i\in\{0,z\}$ can be rescaled to yield MZVs using \eqn{scaling} provided that
the last entry of $\vec{a}$ is different from zero. In a generic situation, the
relation \cite{Broedel:2013tta}
\begin{subequations}
  \label{eqn:GIdentity}
  \begin{align}
  G(a_1,
  \ldots,a_{i-1},z,a_{i+1},\ldots,a_n;z)&=G(a_{i-1},a_1,\ldots,a_{i-1},\hat{z},a_{i+1},
\ldots,a_n;z)\label{line1}\\
  &\qquad-G(a_{i+1},a_1,\ldots,a_{i-1},\hat{z},a_{i+1},\ldots,a_n;z)\label{line2}\\
  &\qquad-\int_0^z\frac{\dd t}{t-a_{i-1}}G(a_1,\ldots,\hat{a}_{i-1},t,a_{i+1},\ldots,a_n;t)
\label{line3}\\
  &\qquad+\int_0^z\frac{\dd t}{t-a_{i+1}}G(a_1,\ldots,a_{i-1},t,\hat{a}_{i+1},\ldots,a_n;t)
\label{line4}\\
  &\qquad+\int_0^z\frac{\dd t}{t-a_1}G(a_2,\ldots,a_{i-1},t,a_{i+1},\ldots,a_n;t)
\label{line5}
  \end{align}
\end{subequations}
allows to recursively remove the argument $z$ from the labels of a multiple
polylogarithm, because the expressions on the right-hand side either have
shorter labels or are free of $z$. A hat denotes the omission of the respective
label, and it is assumed that at least one $a_j \neq 0$. The availability of a
recursive formula like \eqn{eqn:GIdentity} is intrinsic to the moduli space of
Riemann spheres with marked points \cite{Brown:0606}. An explicit discussion
including algorithms is ref.~\cite{Bogner:2014mha}. 

As an identity similar to \eqn{eqn:GIdentity} will be crucial in deriving
relations for \IIEC s in \subsecref{ssec:MEP} below,
let us briefly comment on the application and generalization of
\eqn{eqn:GIdentity}: If the argument $z$ appears multiple times in the label
$\vec{a}$, the first four terms on the right hand side (terms~\eqref{line1} to
\eqref{line4}) have to be evaluated for each occurrence of $z$.  The reduction
will lead to expressions where the labels of the polylogarithms on the right
hand side are independent of $z$ or shorter, which is ensured by cancellations
between neighboring terms.  If $a_n=z$, the term \eqref{line4} has to be
dropped and the term (\ref{line2}) needs to be altered to
$-G(0,a_1,\ldots,a_{i-1},\hat{z};z)$.

Multiple polylogarithms with $a_1=z$ require special attention as well.
However, in order to keep the exposition simple, we will assume that those
polylogarithms have already been taken care of by applying the shuffle
regularization rule \eqn{shufflereg1}.

The following examples (with $a_j\neq z$) are typical relations derived from
the above identity: 
\begin{align}
 G(a_1,0,z;z) &= G(0,0,a_1;z)-G(0,a_1,a_1;z)-G(a_1;z)\z_2\nnl
 G(a_1,z,a_2;z) &= G(a_1,a_1,a_2;z)-G(a_2,0,a_1;z)+G(a_2,a_1,a_1;z)-
G(a_2,a_1,a_2;z)\,.
\end{align}
Proving \eqn{eqn:GIdentity} is straightforward. It relies on writing the
polylogarithm on the left hand side as the integral over its total derivative and
using partial fraction as well as relations \eqref{eqn:derivativeG1} to
\eqref{eqn:derivativeG2} in \appref{app:MPrelations}.  Finally, let us note
that \eqn{eqn:GIdentity} preserves shuffle regularization.  The complete proof
of~\eqn{eqn:GIdentity} as well as numerous examples are contained in section 5 of
ref.~\cite{Broedel:2013tta}. A collection of identities between MZVs can be
found in the multiple zeta value data mine~\cite{Blumlein:2009cf}.

\subsection{Iterated integrals on an elliptic curve}
\label{ssec:MEP}
In this subsection we are going to take a first look at \IIEC s.  In the
following exposition, we will omit several mathematical details, which will be
discussed in \secref{sec:math} below. As {\IIEC}s will turn out to be a
generalization of the multiple polylogarithms discussed above, we will follow
the structure of the previous subsection closely. 

In \eqn{eqn:defG}, the differential $\dd t$ is weighted by
\begin{equation}
  \frac{1}{t-a_i}\,,
  \label{eqn:genus0weight}
\end{equation}
which yields iterated integrals on the genus-zero curve $\ZC \setminus
\{a_1,...,a_n\}$.  Here, we propose a generalization to \IIEC s. 
An infinite number of weighting functions $f^{(n)}$ of weights
$n=0,1,2,\ldots$ is necessary, whose appearance will be justified and whose
precise definition will be provided in \secref{sec:math}. They lead to \IIEC s
in the same way as does \eqn{eqn:genus0weight} at genus zero.  Accordingly, the
functions $f^{(n)}(z,\tau)$ are doubly periodic with respect to the two cycles
of the torus, with modular parameter $\tau$ in the upper half plane  
\begin{equation}
  f^{(n)}(z,\tau)=f^{(n)}(z+1,\tau)\quad\text{and}\quad 
  f^{(n)}(z,\tau)=f^{(n)}(z+\tau,\tau) \ .
  \label{eqn:felliptic}
\end{equation}
Below, we are going to suppress the $\tau$-dependence and will simply write
$f^{(n)}(z)$.  As will be explained in \subsecref{subsec:definition}, the
functions $f^{(n)}$ are known for all non-negative integer weights $n$. In
particular they are non-holomorphic and expressible in terms of the odd Jacobi
function $\theta_1(z,\tau)$, e.g.
\begin{align}
  f^{(0)}(z) &\equiv 1 \ , \ \ \ \ \ \ \ \
  f^{(1)}(z) \equiv \pd \ln \theta_1(z,\tau) + 2\pi i \frac{ \Im z}{\Im \tau}  \label{eqn:f1} 
  \\
  f^{(2)}(z) &\equiv \frac{1}{2} \Big[ \Big( \pd \ln \theta_1(z,\tau) + 
              2\pi i \frac{ \Im z}{\Im \tau} \Big)^2 + 
	      \pd^2  \ln \theta_1(z,\tau) - 
	      \frac{1}{3}{\theta_1'''(0,\tau)  \over   \theta_1'(0,\tau)} \Big]
	      \label{eqn:f2}
\end{align}
where $\pd$ and $'$ denote a derivative in the first argument of $\theta_1$. Their
parity alternates depending on the weight $n$:
\begin{equation}
  f^{(n)}(-z)=(-1)^{n}f^{(n)}(z)\,.
  \label{eqn:fparity}
\end{equation}
The functions $f^{(n)}$ are defined for arbitrary complex arguments $z$.
Restricting to real arguments~$z$, however, will not only simplify
\eqns{eqn:f1}{eqn:f2} but in addition lead to the system of iterated integrals
appropriate for the one-loop open-string calculations in sections~\ref{sec:fourpoint} and
\ref{sec:multipt} below. Hence, in the remainder of the current section, any
argument and label of the \IIEC s to be defined is assumed to
be real. We will comment on the additional ingredients required for generic
complex arguments~$z$ and relate them to multiple elliptic polylogarithms in
\subsecref{subsec:motivation}. 

Employing the functions $f^{(n)}$, \IIEC s are defined in analogy to \eqn{eqn:defG} via
\begin{equation}
  \GLargz{n_1 &n_2 &\ldots &n_r}{a_1 &a_2 &\ldots &a_r} \equiv
  \int^z_0 \dd t \, f^{(n_1)}(t-a_1) \,
  \GLarg{n_2 &\ldots &n_r}{a_2 &\ldots &a_r}{t},
  \label{eqn:defGell}
\end{equation}
where the recursion starts with $\GL(;z) \equiv 1$. Following the terminology
used for $f^{(n)}$ above, the \IIEC{} in \eqn{eqn:defGell} is
said to have \textit{weight} $\sum_{i=1}^r n_i$, and the number $r$ of
integrations will be referred to as its \textit{length}.

The definition of \IIEC s directly
implies a shuffle relation with respect to the combined letters $A_i \equiv
\begin{smallmatrix} n_i  \\ a_i \end{smallmatrix}$ describing the integration
weights $f^{(n_i)}(z-a_i)$,
\begin{equation}
\GL(A_1,A_2,\ldots,A_r;z) \GL(B_1,B_2,\ldots,B_q;z)  = \GL\big(( A_1,A_2,\ldots,A_r) 
\shuffle (B_1,B_2,\ldots,B_q);z\big)\,,
\label{gl2}
\end{equation}
where the shuffle symbol has been defined in \eqn{eqn:shuffle}. Another
immediate consequence of definition \eqref{eqn:defGell} is the
reflection identity
\begin{equation}
      \GLargz{n_1 &n_2 &\ldots &n_r}{a_1 &a_2 &\ldots &a_r}
      =(-1)^{n_1+n_2+\ldots+n_r}
        \GLargz{n_r &\ldots &n_2 &n_1}{z-a_r &\ldots &z-a_2 &z-a_1}\,.
     \label{eqn:reflection}
\end{equation}
Formally reminiscent of the H\"older convolution in \eqn{eqn:Hoelder},
the above reflection identity is valid for all arguments
$z\in\ZC\setminus\{0\}$. It can be proven using the parity properties of the
weighting functions $f^{(n)}$ in \eqn{eqn:fparity} and a reparametrization of
the integration domain. If all the labels $a_i$ vanish, we will often use the
notation
\begin{equation}
  \GL(n_1,n_2,\ldots,n_r;z) \equiv \GLargz{n_1 &n_2 &\ldots &n_r}{0 &0 &\ldots &0}\,.
\label{l2.7a}
\end{equation}

\subsubsection{Elliptic multiple zeta values}
\label{subsec:eMZV}
Evaluating \IIEC s with all $a_i$ equal to $0$ (or
equivalently $a_i=1$ by the periodicity property \eqn{eqn:felliptic}) at $z=1$
gives rise to iterated integrals
\begin{align}
\omega(n_1,n_2,\ldots,n_r) &\equiv \! \! \! \! \! \! \int \limits_{0\leq z_i \leq z_{i+1} \leq 1}  \! \! \! \! \! \!
f^{(n_1)}(z_1) \dd z_1\, f^{(n_2)}(z_2)\dd z_2\, \ldots f^{(n_r)}(z_r) \dd z_r 
\label{l2.7} \\
&= 
\GL(n_r  ,\ldots ,n_2 ,n_1;1) \notag
\end{align}
%\begin{equation}
%\omega(k_1,k_2,\ldots,k_r) \equiv \int_{0\leq z_i \leq z_{i+1} \leq 1} 
%\omega^{(k_1)}(z_1)\omega^{(k_2)}(z_2) \ldots \omega^{(k_r)}(z_r) = 
%\GLarg{k_r &k_{r-1} &\ldots &k_2 &k_1}{0 &0 &\ldots &0 &0}{1}\,,
%\label{l2.7}
%\end{equation}
%
which we will refer to as \textit{elliptic multiple zeta values} or
\textit{eMZV}s for short. They furnish a natural genus-one generalization of
standard MZVs\footnote{In order to distinguish between eMZVs and MZVs, we will
sometimes refer to the latter as \textit{standard} MZVs.} as defined in
\eqn{eqn:defZeta}. The shuffle relation \eqn{gl2} can be straightforwardly
applied to eMZVs
\begin{equation}
\omega(n_1,n_2,\ldots,n_r) \omega(k_1,k_2,\ldots,k_s) = \omega\big( (n_1,n_2,\ldots,n_r) \shuffle (k_1,k_2,\ldots,k_s) 
\big) \ ,
\label{l2.7sh}
\end{equation}
and the parity property \eqn{eqn:fparity} of the functions $f^{(n)}$
implies the reflection identity
\begin{equation}
\omega(n_1,n_{2} ,\ldots ,n_{r-1} ,n_r)=
     (-1)^{n_1+n_2+\ldots+n_r}\omega(n_r,n_{r-1} ,\ldots ,n_2 ,n_1) \ .
\label{l2.7r}
\end{equation}
Note that a similar set of $\omega$'s can be defined by an iterated integral
along the path from $0$ to $\tau$ replacing the integration domain $[0,1]$ in
\eqn{l2.7}. They appear in the modular transformations of eMZVs and naturally
satisfy the properties \eqns{l2.7sh}{l2.7r} as well. Likewise,
the \IIEC s defined in \eqn{eqn:defGell} allow
for a version with integrations on the path from $0$ to $\tau$.

\paragraph{Regularization.} Among the family of functions $f^{(n)}(z)$ used to
define eIIs and eMZVs, only $f^{(1)}(z)$ has a simple pole at zero and its
images under the translations in \eqn{eqn:felliptic}. Therefore, iterated
integrals of the form
\begin{equation}
  \GLargz{n_r &\ldots &n_2 &n_1}{a_r &\ldots &a_2 &a_1}
= \! \! \! \! \! \! 
\int \limits_{0\leq z_i \leq z_{i+1} \leq z}  \! \! \! \! \! \!
f^{(n_1)}(z_1-a_1) \dd z_1\, f^{(n_2)}(z_2-a_2)\dd z_2\, \ldots f^{(n_r)}(z_r-a_r) \dd z_r
\end{equation}
with $n_1=1$ or $n_r=1$ need to be regularized if either $a_1=0$ or $a_r=z$. As
with multiple polylogarithms, the idea is to slightly move the endpoints of the
integration domain by a small parameter, and then to expand in this parameter.
More precisely, one writes the integral
\begin{equation} 
\label{1}
\int \limits_{\varepsilon \leq z_i \leq z_{i+1} \leq z-\varepsilon}  \! \! \! \! \! \!
f^{(n_1)}(z_1-a_1) \dd z_1\, f^{(n_2)}(z_2-a_2)\dd z_2\, \ldots f^{(n_r)}(z_r-a_r) \dd z_r
\end{equation}
as a polynomial in $\ln(-2\pi i \varepsilon)$, where the branch of the
logarithm is chosen such that we have \mbox{$\ln(-i)=-\frac{\pi i}{2}$}. The regularized
value of \eqn{1} is then defined to be the constant term in this expansion.
The additional $-2\pi i$ in the expansion parameter $\ln(-2\pi i \varepsilon)$
ensures that no logarithms appear in the limit $\tau \to i \infty$, and that
eMZVs degenerate to MZVs. A thorough treatment of this degeneration can be
found in ref.~\cite{Enriquez:ellass} and will be exploited in ref.~\cite{WIP}.

\subsubsection{Removing the argument \texorpdfstring{$\boldsymbol{z}$}{z} from the label}
\label{ssec:remove}

As for the multiple polylogarithms, no arguments $z$ are allowed in the
labels $\{a_1\ldots a_r\}$ in order to perform the integration using
\eqn{eqn:defGell}. Therefore we need to find relations, which trade \IIEC s
with one or multiple occurrences of the argument $z$ in the label for \IIEC s
where $z$ appears in the argument exclusively.  The key idea for finding those
relations is to write the \IIEC{} as the integral of its total derivative
\begin{equation}
\GLargz{n_1 &n_2 &\ldots &n_q &\ldots &n_r}{a_1 &a_2 &\ldots &z &\ldots &a_r} =
\int^z_0 \dd t \  \frac{\dd}{\dd t} 
\GLarg{n_1 &n_2 &\ldots &n_q &\ldots &n_r}{a_1 &a_2 &\ldots &t &\ldots &a_r}{t} + 
\lim_{z\rightarrow 0} \GLargz{n_1 &n_2 &\ldots &n_q &\ldots &n_r}{a_1 &a_2 &\ldots &z &\ldots &a_r}  \,.
\label{gl5z}
\end{equation}
This resembles the strategy at genus zero which led to the identity
\eqn{eqn:GIdentity} between multiple polylogarithms. In the
subsequent, we address additional features and subtleties intrinsic to the
elliptic case.  The feasibility of this approach in the elliptic scenario is
discussed in ref.~\cite{BrownLev}, see in particular theorem 26 therein.    
\paragraph{Boundary terms.}
The boundary term at $z=0$ usually drops out from \eqn{gl5z} due to the
vanishing volume of the integration domain. However, the special situation when
all $n_j=1$ leads to the appearance of standard MZVs.  As will be elaborated on
in \secref{sec:math}, the function $f^{(1)}$ is the only source of
singularities in the integration variables, as can be seen from its leading
behavior $f^{(1)}(z)=z^{-1}+ {\cal O}(z)$. Hence, the regime $z\rightarrow 0$
reproduces multiple polylogarithms as defined in \eqn{eqn:defG}:
\begin{align}
\lim_{z\rightarrow 0} \GLargz{1 &1 &\ldots &1}{a_1 &a_2 &\ldots  &a_r} &= 
\lim_{z\rightarrow 0} \int^z_0 \frac{\dd t_1}{t_1-a_1} 
  \int^{t_1}_0 \frac{ \dd t_2}{t_2-a_2} \ldots \int^{t_{r-1}}_0 \frac{ \dd t_r }{t_r-a_r} \nnl
&= \lim_{z\rightarrow 0} G(a_1,a_2,\ldots,a_r;z)\,.
\label{bdy1}
\end{align}
If all $a_j \in \{0,z\}$, the scaling relation \eqn{scaling} allows to rewrite
the polylogarithms in terms of MZVs (see \eqn{eqn:defZeta}), leading to 
\begin{align}
\lim_{z\rightarrow 0} \GLargz{n_1 &n_2 &\ldots &n_r}{b_1 z &b_2 z &\ldots  &b_r z} &= G(b_1,b_2,\ldots,b_r;1) \prod_{j=1}^r \delta_{n_j,1} \ , \ \ \ b_j \in \{0,1\}\,.
\label{bdy2}
\end{align}
\paragraph{Partial derivatives.}
The total $t$-derivative in \eqn{gl5z} can be written in terms of partial
derivatives with respect to the arguments and the labels. This requires the
elliptic analogues of eqns.~\eqref{eqn:derivativeG1} to
\eqref{eqn:derivativeG2} listed below in order to arrive at shorter elliptic
polylogarithms. The derivative with respect to the argument 
\begin{equation}
  \frac{\pd}{\pd z}
  \GLargz{n_1 &n_2 &\ldots &n_r}{a_1 &a_2 &\ldots &a_r}
  = f^{(n_1)}(z-a_1) 
  \GLargz{n_2 &\ldots &n_r}{a_2 &\ldots &a_r}
\label{gl3}
\end{equation}
follows straightforwardly from \eqn{eqn:defGell}. Slightly more work using
$\frac{\pd}{\pd a} f^{(n)}(t-a)=-\frac{\pd}{\pd t} f^{(n)}(t-a)$ as well as
\eqn{gl3} is required for derivatives with respect to labels $a_{q}$. Starting with
the special cases $q=1$ and $q=r$ one finds
\begin{align}
\frac{\pd}{\pd a_1}
&\GLarg{n_1 &n_2 &\ldots &n_r}{a_1 &a_2 &\ldots &a_r}{t_0}
=    - f^{(n_1)}(t_0-a_1)  \GLarg{n_2 &n_3 &\ldots &n_r}{a_2 &a_3 &\ldots &a_r}{t_0} \nnl
&+\int^{t_0}_0 \dd t \, f^{(n_{1})}(t-a_{1}) f^{(n_{2})}(t-
a_{2}) \GLarg{n_3 &\ldots &n_r}{a_3 &\ldots &a_r}{t}
\label{gl11extra}
\\
%%%%%%%%%%%%
\frac{\pd}{\pd a_r}
&\GLarg{n_1 &n_2 &\ldots &n_r}{a_1 &a_2 &\ldots &a_r}{t_0}
=  f^{(n_r)}(-a_r) \GLarg{n_1 &n_2 &\ldots &n_{r-1}}{a_1 &a_2 &\ldots &a_{r-1}}{t_0} \nnl
&- \left( \prod_{j=1}^{r-2} \int^{t_{j-1}}_0 \dd t_j \, f^{(n_j)}(t_j-a_j) \right)   
\int^{t_{r-2}}_0 \dd t \, f^{(n_{r-1})}(t-a_{r-1}) f^{(n_{r})}(t-
a_{r})\,.
\label{gl11a}
\end{align}
Deriving with respect to a label $a_q$ with $q\neq 1,r$  yields
\begin{align}
 & \frac{\pd}{\pd a_q}
  \GLarg{n_1 &n_2 &\ldots &n_r}{  a_1 &a_2 &\ldots &a_r}{t_0}  \notag \\
 &=    \left( \prod_{j=1}^{q-1} \int^{t_{j-1}}_0 \dd t_j \, f^{(n_j)}(t_j-a_j) \right) \int ^{t_{q-1}}_0 \dd t \, f^{(n_q)}(t-a_q) f^{(n_{q+1})}(t-a_{q+1})  \GLarg{n_{q+2} &\ldots &n_r}{ a_{q+2} &\ldots &a_r}{t}    \nnl
 &-  \left( \prod_{j=1}^{q-2} \int^{t_{j-1}}_0 \dd t_j \, f^{(n_j)}(t_j-a_j) \right) \int^{t_{q-2}}_0 \dd t \, f^{(n_{q-1})}(t-a_{q-1}) f^{(n_{q})}(t -a_{q})  \GLarg{n_{q+1} &\ldots &n_r}{ a_{q+1} &\ldots &a_r}{t}.
\label{gl5b}
\end{align}
%
%
%\begin{align}
%  \frac{\pd}{\pd a_q}
%  \GLarg{n_1 &n_2 &\ldots &n_r}{  a_1 &a_2 &\ldots &a_r}{t_0}&= 
% %
%     \left( \prod_{j=1}^{q-1} \int \limits^{t_{j-1}}_0 \dd t_j \, f^{(n_j)}(t_j-a_j) \right)\times\nnl
%     &\ \ \ \ \ \ \times\int \limits^{t_{q-1}}_0 \dd t_q \, f^{(n_q)}(t_q-a_q) f^{(n_{q+1})}(t_q-a_{q+1})  \GLarg{n_{q+2} &\ldots &n_r}{ a_{q+2} &\ldots &a_r}{t_q} \nnl
% %
% &-  \left( \prod_{j=1}^{q-2} \int \limits^{t_{j-1}}_0 \dd t_j \, f^{(n_j)}(t_j-a_j) \right)\times\nnl 
%     &\ \ \ \ \ \ \times\int \limits ^{t_{q-2}}_0 \dd t_{q-1} \, f^{(n_{q-1})}(t_{q-1}-a_{q-1}) f^{(n_{q})}(t_{q-1}-a_{q})  \GLarg{n_{q+1} &\ldots &n_r}{ a_{q+1} &\ldots &a_r}{t_{q-1}} \,.
%\label{gl5b} 
%\end{align}
%

\paragraph{Total derivatives.} Summing the above partial derivatives with
respect to the argument $z$ and the labels $a_q$, total derivatives from
\eqn{gl5z} can be expressed in a very efficient way. For a single instance of
$a_q=z$, the special cases $q=1$ and $q=r$ give rise to
\begin{align}
\frac{\dd}{\dd t_0} 
\GLarg{n_1 &n_2 &\ldots &n_r}{t_0 &a_2 &\ldots  &a_r}{t_0} &= 
\int_0^{t_0} \dd t \, f^{(n_1)}(t-t_0) f^{(n_2)}(t-a_2) 
\GL \left( \begin{smallmatrix}
n_3 &\ldots &n_r \\
a_3 &\ldots &a_r
\end{smallmatrix} ; t\right) \quad\text{and}
\label{tzero1} \\
\frac{\dd}{\dd t_0} 
\GLarg{n_1 &\ldots &n_{r-1} &n_r}{a_1 &\ldots &a_{r-1} &t_0}{t_0} &= 
f^{(n_1)}(t_0-a_1) \GLarg{n_2 &\ldots &n_{r-1} &n_r}{a_2 &\ldots &a_{r-1} &t_0}{t_0} + 
f^{(n_r)}(-t_0) \GLarg{n_1 &\ldots &n_{r-1} }{a_1 &\ldots &a_{r-1} }{t_0}
\nnl
- \Bigg( \prod_{j=1}^{r-2} \int^{t_{j-1}}_0 \dd t_j \; &f^{(n_j)}(t_j-a_j) \Bigg) 
  \int^{t_{r-2}}_0 \dd t \, f^{(n_{r-1})}(t-a_{r-1}) f^{(n_{r})}(t-t_0)  \ . 
\label{tzero2}
\end{align}
For $q\neq 1,r$, the integrand of \eqn{gl5z} takes the form 
\begin{align}
\frac{\dd}{\dd t_0} 
&\GLarg{n_1 &n_2 &\ldots &n_{q-1} &n_q &n_{q+1} &\ldots &n_r}{a_1 &a_2 &\ldots &a_{q-1} &t_0 &a_{q+1} &\ldots &a_r}{t_0} =  
f^{(n_1)}(t_0-a_1) \GLarg{n_2 &\ldots &n_{q-1} &n_q &n_{q+1} &\ldots &n_r}{a_2 &\ldots &a_{q-1} &t_0 &a_{q+1} &\ldots &a_r}{t_0}  \nnl
%%%
+& \left( \prod_{j=1}^{q-1} \int^{t_{j-1}}_0 \dd t_j \, f^{(n_j)}(t_j-a_j) \right) 
\int^{t_{q-1}}_0 \dd t \, f^{(n_q)}(t-t_0) f^{(n_{q+1})}(t-a_{q+1})  
\GLarg{n_{q+2} &\ldots &n_r}{ a_{q+2} &\ldots &a_r}{t}
\label{tzero3}
\\
- &\left( \prod_{j=1}^{q-2} \int^{t_{j-1}}_0 \dd t_j \, f^{(n_j)}(t_j-a_j) \right)
\int^{t_{q-2}}_0 \dd t  \, f^{(n_{q-1})}(t -a_{q-1}) f^{(n_{q})}(t -t_0)  
\GLarg{n_{q+1} &\ldots &n_r}{ a_{q+1} &\ldots &a_r}{t}.
\notag
\end{align}
Further examples with repeated appearances of $t_0$ are displayed in
\appref{ellpolyexpla}.

\paragraph{Fay identities.}
Having applied the above derivative identities, one is usually left with
expressions containing integrals of the form
\begin{equation}
  \int^{z}_0 \dd t\ f^{(n_1)}(t-a_1) f^{(n_2)}(t-a_2)\,,
  \label{}
\end{equation}	
where the integration variable appears in the argument of more than one
function $f^{(n)}$. In the corresponding situation for multiple polylogarithms,
with weights of the form \eqn{eqn:genus0weight}, one would have used partial
fraction identities 
\begin{equation}
  \frac{1}{(t-a)(t-b)}=\frac{1}{(t-a)(a-b)}+\frac{1}{(t-b)(b-a)} 
  \label{eqn:partialfraction}
\end{equation}
in order to avoid the repeated appearance of the integration variable $t$.
Analogous relations for the more general class of weighting functions $f^{(n)}$
are provided by Fay identities, which will be put in a larger mathematical
context in \secref{sec:math} below. They relate products $f^{(n_1)}f^{(n_2)}$
at arguments $x,t$ and $x-t$ and thereby allow to
systematically remove repeated appearances of some integration variable. A
simple example of a Fay identity relates products of functions $f^{(1)}$ to a
sum of functions $f^{(2)}$
\begin{equation}
f^{(1)}(t-x) f^{(1)}(t) = f^{(1)}(t-x) f^{(1)}(x) - f^{(1)}(t) f^{(1)}(x) + 
f^{(2)}(t) + f^{(2)}(x) + f^{(2)}(t-x)  \ .
\label{kron1.34}
\end{equation}
The general relation, which is valid for complex arguments $x,t$ as well, 
\begin{align}
f^{(n_1)}(t-x) f^{(n_2)}(t)
 &=  - (-1)^{n_1} f^{(n_1+n_2)}(x) + \sum_{j=0}^{n_2} { n_1 - 1 + j \choose j} f^{(n_2-j)}(x) f^{(n_1+j)}(t-x) \notag  \\
 & \ \ \ \ \ + \sum_{j=0}^{n_1} {n_2-1+j \choose j} (-1)^{n_1+j} f^{(n_1-j)}(x) f^{(n_2+j)}(t) \ , 
\label{kron1.33b}
\end{align}
in turn allows to remove all repeated occurrences of the variable $t$.
Iterating the above steps, one can thus eliminate all arguments from the label
of any \IIEC{} recursively.

\paragraph{Result.}

Combining the Fay identity \eqn{kron1.33b} with the total derivatives in
eqns.~\eqref{tzero1} to \eqref{tzero3} turns \eqref{gl5z} into a recursive rule
for removing the argument $z$ from the label of $\GLargz{n_1 &\ldots &n_q
&\ldots &n_r}{a_1 &\ldots &z &\ldots &a_r}$. In the equations below, all terms
on the right-hand side are either free of $a_q=z$ or have shorter labels.  The
special cases $q=1$ and $q=r$ yield
\begin{align}
\GLarg{n_1 &n_2 &\ldots &n_r}{z &a_2 &\ldots &a_r}{z} &= 
  \lim_{z\rightarrow 0} G(z,a_2,\ldots,a_r;z) \prod_{j=1}^r \delta_{n_j,1} - 
  (-1)^{n_1} \GLarg{n_1+n_2 &0 &n_3 &\ldots &n_r}{a_2 &0 &a_3 &\ldots &a_r}{z} \nnl
&+\sum_{j=0}^{n_1} (-1)^{n_1+j} {n_2-1+j \choose j}  
  \GLarg{n_1-j &n_2+j &n_3 &\ldots &n_r}{a_2 &a_2 &a_3 &\ldots &a_r}{z}   \nnl
&+\sum_{j=0}^{n_2} {n_1-1+j \choose j} \int^z_0 \dd t \, f^{(n_2-j)}(t-a_2) 
  \GLarg{n_1+j &n_3 &\ldots &n_r}{t &a_3 &\ldots &a_r}{t} \label{extra21} \\
%%%%%%%%%%%%%%%%%%
\GLarg{n_1 &\ldots &n_{r-1} &n_r}{a_1  &\ldots &a_{r-1} &z}{z} &= 
  \lim_{z\rightarrow 0} G(a_1,\ldots,a_{r-1},z;z) \prod_{j=1}^r \delta_{n_j,1} +
  \int^z_0 \dd t \, f^{(n_1)}(t-a_1) \GLarg{n_2 &\ldots &n_{r-1} &n_r}{a_2  &\ldots &a_{r-1} &t}{t}  \nnl
&+ (-1)^{n_r} \GLargz{n_r &n_1 &\ldots &n_{r-1} }{0 &a_1 &\ldots &a_{r-1}} + 
   (-1)^{n_r} \GLargz{n_{r-1}+n_r &n_1 &\ldots &n_{r-2} &0}{a_{r-1} &a_1 &\ldots &a_{r-2} &0} \nnl
&- \sum_{j=0}^{n_{r-1}} {n_r-1+j \choose j} \int^z_0 \dd t \, f^{(n_{r-1}-j)}(t-a_{r-1}) 
   \GLarg{n_1 &\ldots &n_{r-2} &n_r+j}{a_1 &\ldots &a_{r-2} &t}{t} \nnl
&- \sum_{j=0}^{n_r} {n_{r-1}-1+j \choose j} (-1)^{n_r+j} 
   \GLargz{n_r-j &n_1 &\ldots &n_{r-2} &n_{r-1}+j}{a_{r-1} &a_1 &\ldots &a_{r-2} &a_{r-1}}  \ ,
 \label{extra22} 
\end{align}
while $a_q=z$ at a generic position $q\neq 1,r$ can be addressed via
\begin{align}
&\GLargz{n_1 &n_2 &\ldots &n_{q-1} &n_q &n_{q+1} &\ldots &n_r}{a_1 &a_2 &\ldots &a_{q-1} &z &a_{q+1} &\ldots &a_r} = 
  \lim_{z\rightarrow 0} G(a_1,\ldots,a_{q-1},z,a_{q+1},\ldots,a_r;z) \prod_{j=1}^r \delta_{n_j,1}  \nnl
& \ \ + \int^z_0 \dd t \, f^{(n_1)}(t-a_1) \GLarg{n_2 &\ldots &n_{q-1} &n_q &n_{q+1}&\ldots n_r}{a_2 &\ldots &a_{q-1}&t &a_{q+1} &\ldots a_{r}}{t} \nnl
& \ \ - (-1)^{n_q} \GLargz{n_q+n_{q+1} &n_1 &\ldots &n_{q-1} &0 &n_{q+2} &\ldots &n_r}{a_{q+1} &a_1 &\ldots &a_{q-1} &0 &a_{q+2} &\ldots &a_r}
       +(-1)^{n_q} \GLargz{n_q+n_{q-1} &n_1 &\ldots &n_{q-2} &0 &n_{q+1} &\ldots &n_r}{a_{q-1} &a_1 &\ldots &a_{q-2} &0 &a_{q+1} &\ldots &a_r} \nnl
& \ \ + \sum_{j=0}^{n_{q+1}} {n_q-1+j \choose j} \int^z_0 \dd t \, f^{(n_{q+1}-j)}(t-a_{q+1}) 
   \GLarg{n_1 &\ldots &n_{q-1} &n_q+j &n_{q+2} &\ldots &n_r}{a_1 &\ldots &a_{q-1} &t &a_{q+2} &\ldots &a_r}{t}    \nnl
& \ \ + \sum_{j=0}^{n_q} {n_{q+1}-1+j \choose j} (-1)^{n_q+j} 
   \GLargz{n_q-j &n_1 &\ldots &n_{q-1} &n_{q+1}+j &n_{q+2} &\ldots &n_r}{a_{q+1} &a_1 &\ldots &a_{q-1} &a_{q+1} &a_{q+2} &\ldots &a_r} \notag  \\
& \ \ - \sum_{j=0}^{n_{q-1}} {n_q-1+j \choose j} \int^z_0 \dd t \, f^{(n_{q-1}-j)}(t-a_{q-1}) 
   \GLarg{n_1 &\ldots &n_{q-2} &n_q+j &n_{q+1} &\ldots &n_r}{a_1 &\ldots &a_{q-2} &t &a_{q+1} &\ldots &a_r}{t} \nnl
& \ \ - \sum_{j=0}^{n_q} {n_{q-1}-1+j \choose j} (-1)^{n_q+j} 
   \GLargz{n_q-j &n_1 &\ldots &n_{q-2}&n_{q-1}+j &n_{q+1}  &\ldots &n_r}{a_{q-1} &a_1 &\ldots &a_{q-2} &a_{q-1} &a_{q+1} &\ldots &a_r} \label{extra23}
 \,.
\end{align}
Situations with multiple successive appearance of $a_j=z$ are discussed in appendix \ref{ellpolyexpl}.

\paragraph{Examples.}

At length one, 
the reflection identity \eqn{eqn:reflection} implies that
\begin{equation}
\GLargz{n}{z} = (-1)^n \GL (n;z) \,,
\label{lgth1}
\end{equation}
which covers all identities at this length. At length two, cases with $n_1=0$
or $n_2=0$ are similarly determined by \eqn{eqn:reflection}, so the simplest
non-trivial application of \eqn{gl5z} is $\GLargz{1 &1}{z &0}$. The
differential can be derived via \eqn{gl11extra} and simplified using the Fay
identity \eqn{kron1.34} as well as \eqn{lgth1},
\begin{equation}
\frac{\dd }{\dd t}
\GLarg{1 &1}{t &0}{t} =  2 \GL(2;t)+ f^{(2)}(t) \GL(0;t)-2 f^{(1)}(t) \GL(1;t)  \ ,
\label{lgth2}
\end{equation}
see \eqn{l2.7a} for the notation on the right hand side. In combination with the boundary 
term
\begin{equation}
\lim_{z\rightarrow 0} \GLargz{1 &1}{z &0} = G(1,0;1) = \zeta_2 \ ,
\label{lgth3}
\end{equation}
we find
\begin{equation}
\GLargz{1 &1}{z &0} = 2 \GL (0,2;z)+\GL (2,0;z) - 2 \GL(1,1;z) + \zeta_2 \ ,
\label{lgth4}
\end{equation}
which of course agrees with the general formula \eqn{extra21}. The same
reasoning can be applied recursively to obtain for example
\begin{align}
  \GLargz{1 &1 &1}{z &0 &0} &=  
  - \GLargz{1 &1 &1}{z &z &0} =- \GL (
0 ,3 ,0 ; z)  - \GL (0, 0 ,3; z)  -3 \GL (1,1,1;z)+\GL (2 ,0 ,1  ; z)   \nnl
&\ \ \ \ \ \ \ +\GL(1 ,2 ,0 ; z) +2 \GL (
0 ,2 ,1  ; z)+ 2\GL (
1 ,0 ,2 ; z)+ \zeta_2 \GL (1;z) - \zeta_3 
\label{lgth5} \\
%%%%
\GLargz{1 &0 &0 &1}{z &0 &0 &0} &= 2 \GL (0,0,0,2;z)+\GL (0,0,2,0;z) - 2 \GL(0,0,1,1;z) + \zeta_2  \GL(0,0;z) 
\label{lgth5a}
\end{align}
as well as
\begin{align}
\GLargz{0 &1 &0 &1 &0}{0 &z&0 &0&0} &=2 \GL(0,0,0,2,0;z) + \GL(0,2,0,0,0;z) -2 \GL(0,1,0,1,0;z)\label{lgth6}
\\
\GLargz{0 &1 &1 &0 &0}{0 &z&0 &0&0} &=
 \GL(0,0,2,0,0;z) + \GL(0,0,0,2,0;z)+ \GL(2,0,0,0,0;z)\nnl
 & \ \ \ - \GL(1,0,1,0,0;z)- \GL(1,0,0,1,0;z) \ . \label{lgth7}
\end{align}
In \subsecref{sec:stringemzv} these relations turn out to be crucial to
express the low energy expansion of one-loop string amplitudes in terms of
eMZVs.

The most general relation at length two following from \eqn{extra21} reads
\begin{align}
\GL \left( \begin{smallmatrix}
n_1 &n_2  \\
z &0
\end{smallmatrix} ; z\right) &=- 
 (-1)^{n_1} \GL(n_1 + n_2, 0;z) + \sum_{r=0}^{n_2} (-1)^{n_1+r} { n_1-1+r \choose r} \GL(n_2-r,n_1+r;z)
 \notag \\
 & + \sum_{r=0}^{n_1} (-1)^{n_1+r} { n_2-1+r \choose r} \GL(n_1-r,n_2+r;z) + \delta_{n_1,1}\delta_{n_2,1} \zeta_2  \ ,
\label{l2.3a}
\end{align}
and determines $\GLargz{n_1 &n_2}{0 &z}$ through the shuffle identity and
\eqn{lgth1}. Analogous relations at length three can be found in
\appref{lthree}.

\subsubsection{Relations among elliptic multiple zeta values}
\label{sec:eMZVrel}

Apart from their application to string amplitudes, the above manipulations of
\IIEC s are instrumental to derive relations among eMZVs beyond the obvious
reflection and shuffle properties. By definition \eqn{l2.7}, \IIEC s with all
labels $a_j=0$ yield eMZVs in the limit $z\rightarrow 1$ of their argument. At
the level of labels $a_j=z$, the limit $z\rightarrow 1$ is equivalent to
$a_j\rightarrow 0$ since the $f^{(n)}$ are periodic under $z \mapsto z+1$,
hence
\begin{equation}
\lim_{z\rightarrow 1} \GLargz{n_1 &n_2 &\ldots &n_r}{a_1 &a_2 &\ldots &a_r} = \omega(n_r,\ldots,n_2,n_1) \ , \ \ \ a_j \in \{0,z\} \ , \ \ \ n_1,n_r \neq 1 \ .
\label{lim1}
\end{equation}
Note that endpoint divergences caused by the simple pole in $f^{(1)}$ might
introduce additional MZV constants similar to \eqn{bdy2}, that is why the cases
$n_1,n_r=1$ are excluded explicitly. 

At length two, for example, \eqn{l2.3a} implies the following eMZV identity provided that 
the limit $z\rightarrow 1$ is non-singular:
\begin{align}
\omega(n_2,n_1)  &=- 
 (-1)^{n_1} \omega(0,n_1 + n_2) + \sum_{r=0}^{n_2} (-1)^{n_1+r} { n_1-1+r \choose r} \omega(n_1+r,n_2-r)
 \notag \\
 & + \sum_{r=0}^{n_1} (-1)^{n_1+r} { n_2-1+r \choose r} \omega(n_2+r,n_1-r) \ , \ \ \ n_1,n_2\neq 1 \ .
\label{l2.3aa}
\end{align}
At low weights $n_i$, the coefficients in \eqn{l2.3aa} are particularly simple such as
\begin{equation}
\omega(2, 3) = \omega(0, 5)
 \  , \ \ \ \ \ \ 
\omega(3, 4) = -2\,\omega(0, 7) + \omega(2, 5) \ .
\label{l2.8a}
\end{equation}
Similar procedures can be carried out at higher length. Combining
e.g.~\eqn{sumexp2} and a suitable generalization thereof to length four leads to
\begin{align}
0 &= \omega(0, 0, 5) + \omega(0, 1, 4) + \omega(2, 0, 3) \label{l3.21g}\\
0 &= 10 \, \omega(0,0,0,5)+4\,\omega(0,0,3,2)+2\,\omega(0,2,0,3)-\omega(2)\omega(0,3)- \omega(0,5) \ .
\label{l3.21h}
\end{align}
At length five, a combination of
\eqns{lgth6}{lgth7} with the shuffle relation \eqn{l2.7sh} yields
\begin{align}
\omega(0,1,0,1,0) &= \omega(0,2,0,0,0)
\label{l2.8x1}
\\
\omega(0,1,1,0,0) &= \omega(2,0,0,0,0)- \omega(2) \omega(0,0,0,0)  \ ,
\label{l2.8x2} 
\end{align}
which will be applied in \subsecref{sec:stringemzv}.

%%%%%%%%%%%%%%%%%%%%%%%%%%%%%%%%%%%%%%%%%%%%%%%%%%%%%%%%%%%%%%%%%%%%%%%%%%%%%
%%%%%%%%%%%%%%%%%%%%%%%%%%%%%%%%%%%%%%%%%%%%%%%%%%%%%%%%%%%%%%%%%%%%%%%%%%%%%
%%%%%%%%%%%%%%%%%%%%%%%%%%%%%%%%%%%%%%%%%%%%%%%%%%%%%%%%%%%%%%%%%%%%%%%%%%%%%
%%%%%%%%%%%%%%%%%%%%%%%%%%%%%%%%%%%%%%%%%%%%%%%%%%%%%%%%%%%%%%%%%%%%%%%%%%%%%
%%%%%%%%%%%%%%%%%%%%%%%%%%%%%%%%%%%%%%%%%%%%%%%%%%%%%%%%%%%%%%%%%%%%%%%%%%%%%
%%%%%%%%%%%%%%%%%%%%%%%%%%%%%%%%%%%%%%%%%%%%%%%%%%%%%%%%%%%%%%%%%%%%%%%%%%%%%
%%%%%%%%%%%%%%%%%%%%%%%%%%%%%%%%%%%%%%%%%%%%%%%%%%%%%%%%%%%%%%%%%%%%%%%%%%%%%

\section{The functions \texorpdfstring{$\boldsymbol{f^{(n)}}$}{fn} on the elliptic curve}
\label{sec:math}

In this section, we provide the definition and mathematical framework for the
functions $f^{(n)}$, thereby supplementing our heuristic approach in
\secref{sec:polylog}. Before doing so, let us start with some mathematical
motivation, in which we explain in particular why we need -- in distinction to
multiple polylogarithms -- an infinite number of them.

\subsection{Motivation}
\label{subsec:motivation}

The importance of multiple polylogarithms as defined in \eqn{eqn:defG} becomes
evident, when considering \textit{homotopy-invariant} iterated integrals on the
multiply punctured complex plane $\mathbb C \setminus \{a_1,...,a_n\}$: the
value of any such integral evaluated on a path $\gamma$ depends on the homotopy
class of the path only and is a $\mathbb C$-linear combination of multiple
polylogarithms.

Instead of the multiply punctured plane, let us now consider the complex
elliptic curve \mbox{$E_{\tau}=\mathbb C / (\mathbb Z+\mathbb Z\tau)$} with its
origin removed (we write this as $E_{\tau}^{\times}$), where
\mbox{$\Im(\tau)>0$}.  One possible definition of multiple elliptic
polylogarithms is via iterated integrals on $E_{\tau}^{\times}$.  Writing the
canonical coordinate on $E_{\tau}^{\times}$ as $z=s+r\tau$ with $s,r \in
\mathbb R$, such that $r\equiv \frac{ \Im(z)}{\Im(\tau)}$, two natural
differential forms on $E_{\tau}^{\times}$ read
\begin{equation}
  \label{eqn:nu}
  \dd z\quad\text{and}\quad \nu\equiv 2\pi i \, \dd r   \ .
\end{equation}
These differential forms, however, are not sufficient to describe all iterated
integrals on $E^{\times}_{\tau}$.  Even worse, iterated integrals employing the
differential forms $\dd z$ and $\nu$ only will not be homotopy-invariant in
general, i.e.~they will depend on the choice of a path in a given
homotopy class.

Both problems are overcome simultaneously by supplementing \eqn{eqn:nu} by an
infinite tower of differentials $f^{(n)}(z) \dd z$ constructed through a
generating function \cite{BrownLev}\footnote{Note that in ref.~\cite{BrownLev},
$\Omega(z,\alpha,\tau)$ is defined as a differential form, i.e.~includes $\dd
z$.}
\begin{equation}
	\Omega(z,\alpha,\tau)=\sum_{n \geq 0}f^{(n)}(z)\alpha^{n-1}\ ,
	\label{preomega}
\end{equation}
where $f^{(0)}(z)\equiv 1$. In particular, it has been proven in
ref.~\cite{BrownLev} that every iterated integral in $\nu$ and $\dd z$ can be
uniquely lifted to a homotopy-invariant iterated integral over $\nu$ and
$f^{(n)}(z) \dd z$.  Conversely, every homotopy-invariant iterated integral on
$E^{\times}_{\tau}$ arises in this way.

The form of the generating function and its coefficients $f^{(n)}$ in
\eqn{preomega} can be fixed by constructing a doubly-periodic connection $J$
satisfying the integrability condition
\begin{equation}
 \dd J+J \wedge J=0 \ .
  \label{eqn:int}
\end{equation}
This requirement singles out a unique completion of $J=\nu X_0 + \dd z \,
X_1+\ldots$ to a formal power series in non-commuting variables $X_0$ and $X_1$
given by \cite{BrownLev}
\begin{equation}
J=\nu X_0-\ad_{X_0}\Omega(z,-\ad_{X_0},\tau)(X_1)\dd z \ .
\label{intconn}
\end{equation}
It follows from \eqn{eqn:int} that every word in $X_0,X_1$ in the formal power series
\begin{equation}
  \sum_{k=0}^{\infty}\int J^k
  \label{eqn:assoc}
\end{equation}
is a homotopy-invariant iterated integral on $E^{\times}_{\tau}$, and one can
prove that in fact every such iterated integral arises in this way. Therefore,
every homotopy invariant iterated integral on $E^{\times}_{\tau}$ can be
written as a special linear combination of iterated integrals of the
differential forms $f^{(n)}(z)\dd z$ and $\nu$. The differential form $\nu$
\eqn{eqn:nu}, however, vanishes on the real integration path $\gamma(t) \in
\mathbb R$. Hence, the setup in \subsecref{ssec:MEP} based on real variables
leads to elliptic multiple zeta values defined in ref.~\cite{Enriquez:emzv}
without referring to the differential form $\nu$.

Although homotopy invariance is generically lost for the iterated integral over
the forms   $f^{(n_1)} (z_1)\dd z_1\ldots f^{(n_r)}(z_r)\dd z_r$ on the
punctured elliptic curve $E^{\times} _{\tau}$, its value at the real path
$[0,1]$ as in \eqn{l2.7} can in fact be written as a $ \mathbb Z$-linear
combination of coefficients of words in \eqn{eqn:assoc}, again evaluated on the
path $[0,1]$.  In particular, this shows that the eMZVs associated with the
path $[0,1]$ \cite{Enriquez:emzv} are periods of the fundamental group of
$E^{\times}_{\tau}$.

Hence, the \IIEC s defined by \eqn{eqn:defGell} coincide with the elliptic
polylogarithms defined in ref.~\cite{BrownLev} when restricted to the real
line. They can be lifted to honest homotopy-invariant iterated integrals on the
punctured elliptic curve by means of the differential form $\nu$ defined in
\eqn{eqn:nu}. However, generic combinations of $f^{(n)}(z)
\dd z$ accompany several words in $X_0,X_1$ in \eqn{eqn:assoc} and therefore
allow for various homotopy-invariant completions using $\nu$. Iterated
integrals over $\nu$ and $\dd z$, on the other hand, correspond to a single
word in \eqn{eqn:assoc} and therefore have a unique uplift via $f^{(n\geq
1)}(z)\dd z$ towards the elliptic polylogarithms of ref.~\cite{BrownLev}.

\subsection{Doubly-periodic functions and generating series}

In this section, we define the functions $f^{(n)}$ through a generating series,
closely following ref.~\cite{BrownLev}. In the sequel, $z$ and $\alpha$ are
complex coordinates on $E_\tau^{\times}$. Simultaneously, $\alpha$ will be used
as a formal expansion variable below.  The modular parameter often appears in
the combination
\begin{equation}
%	v=e^{2\pi iz} \hspace{1em}
%	w=e^{2\pi i\alpha} \hspace{1em}
	q\equiv e^{2\pi i\tau} \ ,
	\label{eq:q}
\end{equation}
where $\Im(\tau)>0$ translates into $|q|<1$, relevant for convergence issues.

\subsubsection{Some doubly-periodic functions} \label{sssec:ellf}
A general reference on doubly-periodic functions is ref.~\cite{Weil:1976}.
Let $\theta_1$ denote the odd Jacobi function\footnote{The subsequent
definitions of $f^{(n)}$ are unchanged by $z$-independent rescalings of
$\theta_1$. Hence, the current setup is consistent with
refs.~\cite{BrownLev,LevinRacinet}, which rely on
$\theta(z,\tau)\equiv 2iq^{1/12}\sin(\pi z)\prod_{j=1}^{\infty}(1-e^{2\pi i
z}q^j)\prod_{j=1}^{\infty}(1-e^{-2\pi i z}q^j)$.} defined by
\begin{equation}
	\theta_1(z,\tau)\equiv 2iq^{1/8}\sin(\pi z)\prod_{j=1}^{\infty}(1-q^j)\prod_{j=1}^{\infty}(1-e^{2\pi i z}q^j)\prod_{j=1}^{\infty}(1-e^{-2\pi i z}q^j)\ ,
	\label{thet}
\end{equation}
subject to the following periodicity properties
\begin{equation}
\theta_1(z+1,\tau)=-\theta_1(z,\tau) \ , \ \ \ \ \ \ 
\theta_1(z+\tau,\tau)=-e^{-\pi i\tau}e^{-2\pi iz}\theta_1(z,\tau) \ .
\label{periodicity}
\end{equation}
For $j\geq 1$ we also define the \emph{Eisenstein function} $E_j(z,\tau)$ and
the \emph{Eisenstein series} $e_j(\tau)$ by\footnote{The two cases $j=1,2$ require
the Eisenstein summation prescription
\[
	\sum_{m,n \in \mathbb Z}a_{m,n}\equiv
	\lim_{N \to \infty}\lim_{M \to \infty} \sum_{n=-N}^N\sum_{m=-M}^Ma_{m,n} \ .
\]
}
\begin{equation}
	E_j(z,\tau)\equiv \sum_{m,n \in \mathbb Z}\frac{1}{(z+m+n\tau)^j} \hspace{2em}
	e_j(\tau)\equiv \sum\limits_{m,n \in \mathbb Z \atop{(m,n) \neq (0,0)}} \frac{1}{(m+n\tau)^j}
	\label{eiss}
\end{equation}
which are related to the function $\theta_1(z,\tau)$ via 
\begin{equation}
\frac{\pd}{\pd z}\ln(\theta_1(z,\tau))=E_1(z,\tau) \ ,
\hspace{2em} 
\frac{\pd}{\pd z}E_j(z,\tau)=-jE_{j+1}(z,\tau) 
	 \ .
	 \label{alt0}
\end{equation}

\subsubsection{The Eisenstein-Kronecker series}

The Eisenstein-Kronecker series $F(z,\alpha,\tau)$ is defined by
\cite{Kronecker,BrownLev}
\begin{equation}
F(z,\alpha,\tau) \equiv 
\frac{\theta_1'(0,\tau)\theta_1(z+\alpha,\tau)}{\theta_1(z,\tau)\theta_1(\alpha,\tau)} \ ,
\label{alt1}
\end{equation}
where $'$ denotes a derivative with respect to the first argument. Taking the
logarithmic derivative of \eqn{alt1} together with the Taylor expansion
$E_1(\alpha,\tau)=\frac{1}{\alpha} - \sum_{j=0}^{\infty} \alpha^j
e_{j+1}(\tau)$ leads to the following alternative representation
\cite{ZagierF,LevinRacinet}
\begin{equation}
F(z,\alpha,\tau)=
\frac{1}{\alpha}\exp\left(-\sum_{j\geq 1}\frac{(-\alpha)^j}{j}(E_j(z,\tau)-e_j(\tau))\right)
\label{alt2}
\end{equation}
in terms of the Eisenstein functions and Eisenstein series defined in
\eqn{eiss}. The periodicity properties of the $\theta_1$-function in
\eqn{periodicity} imply that the Eisenstein-Kronecker series is quasi-periodic,
\begin{equation}
				F(z+1,\alpha,\tau)=F(z,\alpha,\tau),\hspace{2em}
				F(z+\tau,\alpha,\tau)=e^{-2\pi i \alpha}F(z,\alpha,\tau) \ .
				\label{alt3}
\end{equation}
Moreover, the representation (\ref{alt2}) together with the Fay trisecant
equation \cite{mumford1984tata} yields the Fay identity
\begin{align}
				F(z_1,\alpha_1,\tau)F(z_2,\alpha_2,\tau)&=F(z_1,\alpha_1+\alpha_2,\tau)F(z_2-z_1,\alpha_2,\tau) \notag
				\\
				&+F(z_2,\alpha_1+\alpha_2,\tau)F(z_1-z_2,\alpha_1,\tau) \ .
				\label{alt4}
\end{align}

\subsubsection{Restoring double periodicity and modularity}

The quasi-periodicity of the Eisenstein-Kronecker series under $z\rightarrow
z+\tau$ as given in \eqn{alt3} can be lifted to an honest periodic
behavior by defining
\begin{equation}
\Omega(z,\alpha,\tau)\equiv 
\exp \bigg( 2\pi i \alpha
\frac{\Im(z)}{\Im(\tau)} \bigg) F(z,\alpha,\tau)\,.
\label{alt5}
\end{equation}
Clearly, the resulting function $\Omega(z,\alpha,\tau)$ is doubly-periodic in $z$, 
\begin{equation}
\Omega(z+1,\alpha,\tau)=
				\Omega(z+\tau,\alpha,\tau)=\Omega(z,\alpha,\tau) \ ,
				\label{alt6}
\end{equation}
and holomorphicity of the Eisenstein-Kronecker series \eqn{alt1} gives rise to
the differential equation
\begin{equation}
	\frac{\partial }{\partial \bar z} \Omega(z,\alpha,\tau)=-\frac{\pi \alpha}{\Im(\tau)}\Omega(z,\alpha,\tau)\, .
	\label{alt7}
\end{equation}
%Ä
The latter implies that the connection $J$ in \eqn{intconn} satisfies the 
integrability condition \eqn{eqn:int} and generates homotopy-invariant iterated integrals 
via the formal power series \eqn{eqn:assoc} \cite{BrownLev}.

Upon taking the exponential in \eqn{alt5} into account, the modular
transformation properties of the Eisenstein-Kronecker series \cite{ZagierF,Hain},
can be translated into
\begin{equation}
\Omega \bigg(\frac{z}{c\tau +d},\frac{\alpha}{c\tau+d},\frac{a\tau+b}{c\tau+d} \bigg) =
(c\tau+d) \Omega(z,\alpha,\tau)  
\label{alt8}
\end{equation}
for $\left( \begin{smallmatrix} a &b \\ c &d \end{smallmatrix} \right) \in
  SL(2,\mathbb Z)$. The Fay identity \eqn{alt4} for the Eisenstein-Kronecker series
  carries over to
\begin{align}
				 \Omega(z_1,\alpha_1,\tau) \Omega(z_2,\alpha_2,\tau)&= \Omega(z_1,\alpha_1+\alpha_2,\tau) \Omega(z_2-z_1,\alpha_2,\tau) \notag
				 \\
				&+ \Omega(z_2,\alpha_1+\alpha_2,\tau) \Omega(z_1-z_2,\alpha_1,\tau)\,
				\label{alt9}
\end{align}
after multiplication with
$\exp \big( \frac{2\pi i}{\Im (\tau)} \big[\alpha_1 \Im(z_1)+\alpha_2
\Im(z_2)\big] \big)$.

\subsection{Definition and properties of the weighting functions
  \texorpdfstring{$\boldsymbol{f^{(n)}}$}{fn}}
\label{subsec:definition}

\subsubsection{Definition of \texorpdfstring{$\boldsymbol{f^{(n)}}$}{fn}}

We define the functions $f^{(n)}$ entering the \IIEC s \eqn{eqn:defGell}
through the following Taylor series in~$\alpha$,
\begin{equation}
\alpha \Omega(z,\alpha,\tau) \equiv \sum_{n=0}^{\infty}f^{(n)}(z,\tau)\alpha^{n} 
\label{alt10} \ .
\end{equation}
They are real analytic on the punctured elliptic curve $E^{\times}_{\tau}$. As
above, we will omit the argument $\tau$ and write $f^{(n)}(z)$ or often simply
$f^{(n)}$. Their explicit form is conveniently captured by the following
functions\footnote{Note that all $\cE_n$ are meromorphic except
for $\cE_1$ (due to the term $\Im(z)$), and that $\cE_2(z) = -\wp(z)$ is the
Weierstrass function. Higher functions $\cE_n$ at $n\geq 3$ are related to
derivatives of the Weierstrass function, e.g.  $\cE_3=-\frac{1}{2} \partial
\wp$ and $\cE_4=e_4-\frac{1}{6} \partial^2 \wp$.} ${\cal E}_n$ 
\begin{equation}
{\cal E}_1(z,\tau) \equiv E_1(z,\tau) + 2\pi i \frac{ \Im(z)}{\Im(\tau)} \ , \ \ \ \ \ \ {\cal E}_n(z,\tau) \equiv(-1)^n \big( e_n(\tau) - E_n(z,\tau)\big) \ \forall \ n\ge 2  \ .
\label{calee}
\end{equation}
These functions result in a simple representation of the generating series
\begin{equation}
\alpha \Omega(z,\alpha,\tau) = \exp \left( \sum_{j=1}^{\infty} \frac{\alpha^j}{j} {\cal E}_j(z,\tau) \right)\,,
\label{calee2}
\end{equation}
and allow for a combinatorial interpretation of $f^{(n)}(z,\tau)$ in terms
of the cycle index of the symmetric group $S_n$ (see \appref{app:cyc}).

Comparison with \eqn{alt10} yields the following expressions for the lowest
functions $f^{(n)}$
\begin{align}
\f1 &= \cE_1 \notag \\
\f2 &= {1\over 2}\big(\cE_1^2 + \cE_2\big)\notag \\
\f3 &= {1\over 3!}\big(\cE_1^3 + 3 \cE_1\cE_2 + 2\cE_3\big) \label{fns} \\
\f4 &= {1\over 4!}\big(\cE_1^4 + 6 \cE_1^2\cE_2 + 8\cE_1\cE_3 + 3\cE_2^2 + 6\cE_4\big)\notag \\
\f5 &= {1\over 5!}\big(\cE_1^5 + 10 \cE_1^3\cE_2 + 20\cE_1^2\cE_3 + 15\cE_1\cE_2^2 +
30\cE_1\cE_4+20\cE_2\cE_3 + 24\cE_5\big)  \ . \notag 
\end{align}
The functions ${\cal E}_j$ can be expressed in terms of $\ln \theta_1$ via
\eqn{alt0}, which leads to the representations for $f^{(1)}$ and $f^{(2)}$ provided in
\eqns{eqn:f1}{eqn:f2}. As shown in \appref{app:cyc}, the general expression for $f^{(n)}$ following from \eqn{calee2} reads
\begin{equation}
f^{(n)}= \sum_{a_1,a_2,\ldots,a_n\geq 0} \delta\left(\sum_{i=1}^n i a_i - n\right)\prod_{j=1}^{n} \frac{ {\cal E}_j^{a_j} }{j^{a_j} a_j!} \ ,
\label{alt21}
\end{equation}
and an equivalent recursive representation is given by
\begin{equation}
f^{(n)}= \frac{1}{n} \sum_{j=1}^{n} {\cal E}_j f^{(n-j)}\,.
\label{alt22}
\end{equation}

\subsubsection{Properties of \texorpdfstring{$\boldsymbol{f^{(n)}}$}{fn}}

The functions $f^{(n)}$ inherit their double periodicity, the form of their
antiholomorphic derivative as well as their behavior under modular
transformations from the generating series in eqns.~\eqref{alt6}, \eqref{alt7}
and \eqref{alt8}:
\begin{align}
f^{(n)}(z+1)&=
				f^{(n)}(z+\tau)=f^{(n)}(z)
\label{alt11} \\
\frac{ \partial f^{(n)}(z)}{\partial \bar z}&=
				-\frac{\pi}{\Im (\tau)} f^{(n-1)}(z)
\label{alt12} \\
f^{(n)}\Big(\frac{z}{c\tau+d},\frac{a\tau+b}{c\tau+d} \Big)&=
				(c\tau+d)^n f^{(n)}(z,\tau) \ .
\label{alt13}
\end{align}
Likewise, the Fay identity \eqn{alt9} implies for $f^{(n)}_{ij} \equiv
f^{(n)}(z_i-z_j)$:
\begin{align}
f^{(m-1)}_{il} f^{(n)}_{jl}+ f^{(m)}_{il} f^{(n-1)}_{jl} =  \sum_{r=0}^n {m-1+r\choose r} f^{(n-r)}_{ji} f^{(m-1+r)}_{il}
+ \sum_{r=0}^m {n-1+r\choose r} f^{(m-r)}_{ij} f^{(n-1+r)}_{jl} \ . \label{kron1.33a} 
\end{align}
This identity has been used repeatedly to derive relations among \IIEC s in
\secref{sec:polylog} (cf. \eqn{kron1.33b} above).

Given the singular factor $\frac{
\theta_1'(0,\tau)}{\theta_1(z,\tau)}= \frac{1}{z}+ {\cal O}(z)$ in the Eisenstein-Kronecker
series \eqn{alt1}, one can check that the residue at the simple pole of $\Omega$ at
the origin is independent on $\alpha$. Hence, only $f^{(1)}$ has a simple pole
at any $z=k+\tau l$ for $k,l\in \mathbb Z$
whereas all other weighting functions $f^{(n\neq 1)}$ are regular on the entire
elliptic curve:
\begin{equation}
\lim_{z\rightarrow 0} z f^{(n)}(z) = \delta_{n,1} \ .
\label{alt15}
\end{equation}
It is this property of the functions $f^{(n)}$, which is responsible for the
$z\rightarrow 0$ behavior stated in \eqn{bdy2}.

\subsubsection{\texorpdfstring{$\boldsymbol{q}$}{q}-expansions of \texorpdfstring{$\boldsymbol{f^{(n)}}$}{fn}}
\label{sec:gf}

The Eisenstein-Kronecker series \eqn{alt1} is known to have the following power-series
expansion in $q=e^{2\pi i\tau}$ \cite{Weil:1976,BrownLev} 
\begin{align}
\alpha F(z,\alpha,\tau) &= 1 + \pi \alpha \cot(\pi z) - 2 \sum_{k=1}^{\infty} \zeta_{2k} \alpha^{2k}
 -2\pi i \alpha \! \! \sum_{m,n=1}^{\infty}  \! \! \Big(e^{2\pi i (mz+n\alpha)}-e^{-2\pi i (mz+n\alpha)}\Big) q^{mn} \notag \\
 &\equiv \sum_{n=0}^{\infty} g^{(n)}(z) \alpha^n
  \ .
 \label{kron7.60}
\end{align}
Disentangling the powers of $\alpha$ yields the holomorphic parts $g^{(n)}$ of
the functions $f^{(n)}$, e.g.
\begin{align}
g^{(1)}(z) &= \pi  \cot(\pi z) + 4\pi \sum_{m=1}^{\infty}  \sin(2\pi m z) \sum_{n=1}^{\infty} q^{mn}
\label{kron7.6} \\
g^{(2)}(z) &= -2 \zeta_2 + 8\pi^2 \sum_{m=1}^{\infty}  \cos(2\pi m z) \sum_{n=1}^{\infty} n q^{mn} 
\label{kron7.6a} \\
g^{(3)}(z) &=- 8\pi^3 \sum_{m=1}^{\infty}  \sin(2\pi m z) \sum_{n=1}^{\infty}n^2 q^{mn} \ ,
\label{kron7.8}
\end{align}
where $\cot(\pi z)=\frac{1}{\pi z}+ {\cal O}(z)$ captures the simple pole of $f^{(1)}$. More generally, we find
\begin{align}
g^{(k)}(z) \Big|_{k=2,4,\ldots} &= - 2 \Big[ \zeta_k + \frac{ (2\pi i)^k }{(k-1)!} \sum_{m=1}^{\infty}  \cos(2\pi m z) \sum_{n=1}^{\infty}n^{k-1} q^{mn} \Big]
\label{kron7.10} \\
g^{(k)}(z) \Big|_{k=3,5,\ldots}&= - 2i \frac{ (2\pi i)^k }{(k-1)!} \sum_{m=1}^{\infty}  \sin(2\pi m z) \sum_{n=1}^{\infty}n^{k-1} q^{mn}  \ .
\label{kron7.11} 
\end{align}
The non-holomorphic piece in $f^{(n)}$ consisting of factors $\frac{ \Im(z)
}{\Im(\tau)}$ can be immediately restored via
\begin{equation}
f^{(n)}(z) = \sum_{k=0}^n \frac{ \big[2\pi i \Im (z) \big]^k}{k! \big[ \Im(\tau) \big]^k}
g^{(n-k)}(z) \ .
\label{kron7.14}
\end{equation}
Even though the functions $f^{(n)}$ in the definition \eqn{eqn:defGell} of \IIEC s
are evaluated at real arguments in the subsequent, we will keep 
track of the admixtures of $\Im(z)$ in \eqn{kron7.14} for further applications beyond
this work. For example, another system of \IIEC s and eMZVs can
be defined for the path from $0$ to $\tau$ instead of the real interval $[0,1]$
whose properties are crucially affected by the factors of $\Im (z) $ and the
resulting modular properties.

\section{The one-loop four-point amplitude in open string theory}
\label{sec:fourpoint}

Iterated integrals defined on an elliptic curve in \subsecref{ssec:MEP}
appear naturally in superstring theory. Calculating one-loop scattering
amplitudes among open string states amounts to evaluating iterated
integrals weighted by the functions $f^{(n)}$ defined in \secref{sec:math}.
Accordingly, the expansion of one-loop superstring amplitudes in the inverse
string tension $\ap$ involves eMZVs. 

The $\ap$-expansion of tree-level amplitudes in open string theory is
well known to involve standard MZVs, see e.g.~ref.~\cite{Oprisa:2005wu}. The pattern
of their appearance is much simpler as compared to the MZVs and polylogarithms
in loop amplitudes of field theory and can be understood in terms of motivic
MZVs \cite{Schlotterer:2012ny} as well as the Drinfeld associator
\cite{Broedel:2013aza}. 
Hence, it is not surprising that one-loop string amplitudes furnish a perfect
laboratory to study patterns and properties of eMZVs.  

\begin{figure}
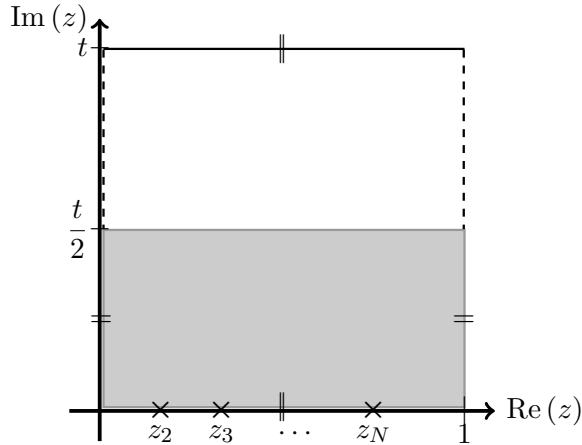

\begin{center}
\tikzpicture[line width=0.5mm]
\draw[->] (-0.4,0) -- (5.2,0)node[right]{$\Re(z)$};
\draw[->] (0,-0.4) -- (0,5.2)node[left]{$\Im(z)$};
\draw[dashed, line width=0.3mm] (0.05,4.79) -- (0.05,2.41);
\draw[dashed, line width=0.3mm] (4.79,4.79) -- (4.79,2.41);
\draw[line width=0.3mm](0.05,4.8) -- (4.8,4.8);
\draw[fill=gray, opacity=0.4,line width=0.3mm] (0.05,0.05) rectangle (4.8,2.4);
\draw (0.8,0.02)node{$\boldsymbol{\times}$} ;
\draw (1.6,0.02)node{$\boldsymbol{\times}$} ;
\draw (3.6,0.02)node{$\boldsymbol{\times}$} ;
\draw (0.8,-0.3)node{$z_2$} ;
\draw (1.6,-0.3)node{$z_3$} ;
\draw (3.6,-0.3)node{$z_N$} ;
\draw (2.6,-0.3)node{$\cdots$} ;
%
%\draw (1.2,2.4)node{$\times$} node[above]{$y_1$};
%\draw (2.2,2.4)node{$\times$} node[above]{$y_2$};
%\draw (4,2.4)node{$\times$} node[above]{$y_M$};
%\draw (3.1,2.4)node{$ $} node[above]{$\cdots$};
%
\draw (4.8,0)node{$|$};
\draw (4.8,-0.3)node{$1$};
\draw (0,2.41)node{$-$}node[left]{$\displaystyle \frac{t}{2}$};
\draw (0,4.82)node{$-$}node[left]{$\displaystyle t$};
\draw (0.02,1.2) node{$=$};
\draw (4.79,1.2) node{$=$};
\draw (2.4,0.05) node{$| \! |$};
\draw (2.4,4.8) node{$| \! |$};
\endtikzpicture
\caption{Parametrization of the cylinder worldsheet through the shaded region.
The boundary under investigation has real coordinates $z_j \in [0,1]$.
The identified edges inherited from the underlying
torus at $\tau=it$ are marked by $=$ and $| \!  |$,
respectively.} 
\label{fig:cyl}
\end{center}
\end{figure}

Iterated integrals in one-loop open string amplitudes occur on the boundaries
of a two-dimensional worldsheet of either cylinder or M\"obius-strip topology \cite{Green:1987mn}. They describe conformally inequivalent configurations of inserting open string
states on the respective boundaries. As a first field of application for eMZVs, we will entirely focus on cylindrical worldsheets in
this work with all integrations confined to one boundary\footnote{The interplay
  between open string worldsheets of different topologies is crucial for the
  cancellations of infinities \cite{Green:1984ed} and anomalies
  \cite{Green:1984sg, Green:1984qs} which occur for gauge group $SO(32)$.}. As 
  shown in figure \ref{fig:cyl}, this situation can be described by a torus with purely 
  imaginary  modular parameter $\tau=it$ with $t\in \mathbb R$. The cylinder boundaries 
  are then parametrized by $\Re(z_j) \in [0,1]$ with $\Im(z_j)=0$ and $\Im(z_j)=\frac{t}{2}$, 
  respectively. The configuration of interest with one boundary empty is
captured by real insertion points $z_j \in \mathbb R$.

\subsection{The four-point amplitude}

For massless open-string excitations in ten dimensions -- gluons and gluinos -- supersymmetry requires
at least four external states for a non-vanishing one-loop amplitude, so the
simplest case to be studied below is the four-point function
\cite{Green:1981ya, Schwarz:1982jn},
\begin{align}
A_{\te{string}}^{\te{1-loop}}(1,2,3,4) &=   s_{12}s_{23} A_{\te{YM}}^{\te{tree}}(1,2,3,4) \int_0^{\infty} \dd t \ I_{4\te{pt}}(1,2,3,4)
\label{4pta} \\
 I_{4\te{pt}}(1,2,3,4)&\equiv  \int^{1}_0 \dd z_4 \int^{z_4}_0 \dd z_3 \int^{z_3}_0 \dd z_2 \int^{z_2}_0 \dd z_1  \,\delta(z_1) \prod_{j<k}^4 \exp\big[ s_{jk} P_{jk} \big]  \ .
\label{4ptaa} 
\end{align}
The entire polarization dependence is captured by the
four-point tree amplitude of sYM field theory, see \cite{Green:1987sp} for its tensor structure.
The worldsheet integral
$I_{4\te{pt}}(1,2,3,4)$ depends on the external momenta $k_i$ through
dimensionless Mandelstam invariants
\begin{equation}
s_{ij} \equiv \ap (k_i+k_j)^2 \ ,
\label{4ptb}
\end{equation}
where momentum conservation and the mass-shell condition $k_i^2=0$ leave two
independent $s_{ij}$,
\begin{equation}
s_{34}=s_{12} \ , \ \ \ \ \ \ s_{14} = s_{23}  \ , \ \ \ \ \ \ s_{13}=s_{24}=-s_{12}-s_{23} \ .
\label{4ptc}
\end{equation}
The dependence on worldsheet positions $z_j \in [0,1]$ enters through the
genus-one Green function
\begin{equation}
  P(z_i-z_j) \equiv \frac{1}{2}  \ln \left| \frac{ \theta_1(z_i-z_j,\tau)}{\theta_1'(0,\tau)}  \right|^2 - \frac{\pi}{\Im(\tau)} \big[ \Im(z_i-z_j) \big]^2 \ .
\label{4ptd}
\end{equation}
which is related to the singular function $f^{(1)}(z_i-z_j)$ in \eqn{eqn:f1} via
%%
%\begin{equation}
%  \pd_{z_i} P(z_i-z_j) = \frac{1}{2}\,f^{(1)}(z_i-z_j)  \,.
%%P_{ij} = \int^{z_i}_{z_j} \dd w \, f^{(1)}(w-z_j) \ .
%\label{4ptda}
%\end{equation}
%%
%
\begin{equation}
  \pd_{z} P(z) = \frac{1}{2}\,f^{(1)}(z)  \,.
%P_{ij} = \int^{z_i}_{z_j} \dd w \, f^{(1)}(w-z_j) \ .
\label{4ptda}
\end{equation}
The non-holomorphic piece in 
%$f^{(1)}(z) \equiv \pd_z \ln \theta_1(z,\tau) + 2\pi i \frac{ \Im z}{\Im \tau}$ 
$f^{(1)}(z) \equiv \frac{\theta_1'(z,\tau)}{\theta_1(z,\tau)} + 2\pi i \frac{ \Im z}{\Im \tau}$ 
 drops out for the
present cylinder parametrization where all vertices are inserted on the
boundary with 
%real coordinates~$x_j$.
real coordinate~$x$. Furthermore, reality of all insertion
points allows to rewrite \eqn{4ptda} as
\begin{equation}
  \pd_{x} P(x) = f^{(1)}(x)  \,.
\label{4ptdb}
\end{equation}
%
%%
%\begin{equation}
%  \pd_{x_i} P(x_i-x_j) = f^{(1)}(x_i-x_j)  \,.
%\label{4ptdb}
%\end{equation}
%%
where we have used $\pd_z=\frac{1}{2}(\pd_x-i\pd_y)$ and $z=x+iy$
with $x,y \in \ZR$.
Integrating along a real path, we find 
\begin{equation}
  P_{ij} = P(x_i-x_j)=\int^{x_i}_{x_j} \dd u \, f^{(1)}(u-x_j) =\int_0^{x_i-x_j} \dd v f^{(1)}(v)\ .
\end{equation}
The endpoint divergence as $u \to x_j$ can be dealt with
through the regularization prescription \eqn{1} which heuristically amounts to
$\lim_{x_i\rightarrow x_j} P_{ij}=0$. Note that the dependence of
$I_{4\te{pt}}(1,2,3,4)$ on $s_{ij}$ and $q\equiv e^{-2\pi t}$ is suppressed for
ease of notation. 

Accordingly, the differential form $\nu
\sim \dd\,\Im(z)$ in \eqn{eqn:nu} required for homotopy invariance does not
contribute to the cylinder integrals under consideration. However, the
admixtures of $\frac{ \Im z}{\Im \tau}$ in $f^{(n)}$ are crucial for modular
invariance of closed-string amplitudes and cylinder diagrams with open string
states on both boundaries.

Translation invariance on genus-one surfaces can be used to fix $z_1=0$.
In addition, the $N$-point integration measure which appears for $N=4$ in
\eqn{4ptaa},
\begin{equation}
\int_{12\ldots N} \equiv \int_{0}^1 \dd z_N \int^{z_N}_0 \dd z_{N-1}  \ldots \int^{z_3}_0 \dd z_2 \int^{z_2}_0 \dd z_1 \, \de(z_1)  \ ,
\label{4pte}
\end{equation}
is invariant under cyclic shifts $z_i \rightarrow z_{i+1 \, \te{mod} \, N}$
and, up to a sign $ (-1)^N$, under reflection $z_i \rightarrow z_{N+1-i}$. Some
features of the one-loop $N$-point amplitudes are discussed in section
\ref{sec:multipt}. Their integrand then involves factors of
$f^{(w_i)}(z_j-z_k)$ with overall weight $\sum_{i} w_i=N-4$.

As another generalization
of the one-loop amplitude \eqn{4pta} in ten spacetime dimensions, one could
consider supersymmetry-preserving compactifications on a torus.  For each
circular dimension of radius $R$, the associated momentum components are
quantized and contribute a 
correction factor of $\sum_{n=-\infty}^{\infty} e^{- n^2 \pi t  R^2/\ap}$ to
the $t$-integrand \cite{Green:1982sw}. Since this does not affect the 
$z_j$-integrations within $I_{4\te{pt}}(1,2,3,4)$ and the resulting eMZVs, the
subsequent results on the $\ap$-expansion are universal for any torus
compactification to spacetime dimensions $D\leq 10$.

\subsection{The \texorpdfstring{$\boldsymbol{\alpha'}$}{a}-expansion}

In this section, we investigate the $\ap$-expansion of the $t$-integrand in
\eqn{4pta},
\begin{align}
 I_{4\te{pt}}(1,2,3,4)&= \int_{1234} \prod_{i<j}^4 \sum_{n_{ij}=0}^{\infty} \frac{ 1}{n_{ij}!} (s_{ij} P_{ij})^{n_{ij}} \ ,
 \label{4ptf}
 \end{align}
which encodes the low-energy effective action for the gluon supermultiplet.
Expanding in $\ap$ amounts to Taylor expanding the exponential in \eqn{4ptaa}
in all the Mandelstam invariants $s_{ij}$ defined in \eqn{4ptb} as well as the
corresponding worldsheet Green function $P_{ij}$ given by \eqn{4ptda}. 
  
In addition to the power-series expansion in $\ap$ discussed in the subsequent,
the integration region of large $t$ in the amplitude \eqn{4pta} gives rise to
logarithmic, non-analytic momentum dependence. The associated threshold
singularities in $s_{ij}$ are for instance crucial to make contact with the
Feynman box integral in the sYM amplitude arising in the point-particle limit
\cite{Green:1982sw}. Mimicking the low energy-analysis of closed string
one-loop amplitudes \cite{Green:1999pv, Green:2008uj, Richards:2008jg, Green:2013bza}, we separate the analytic from the non-analytic
parts of the amplitude and do not keep track of the non-analytic threshold
singularities.  
  
The simplest monomials in $P_{ij}$ inequivalent under cyclic shifts and
reflections of the vertex positions $z_j$ integrate to
\begin{equation}
c_0 \equiv \int_{1234} 1 \ , \ \ \ \ \ \ \ \ \ 
c_1^1 \equiv \int_{1234} P_{12}  \ , \ \ \ \ \ \ \ \
c_2^1\equiv \int_{1234} P_{13}  \ .
\label{4ptg}
\end{equation}
At second and third order in $\ap$ one finds
\begin{align}
c_1^2 &\equiv \frac{1}{2} \int_{1234}  P_{12}^2 \co c^2_3 \equiv \int_{1234} P_{12} P_{14} 
\co c^2_5  \equiv \int_{1234} P_{12} P_{34}  \nnl
c_2^2 &\equiv \frac{1}{2} \int_{1234}P_{13}^2 \co c^2_4 \equiv \int_{1234} P_{13} P_{24} 
\co c^2_6  \equiv \int_{1234} P_{12} P_{13}\label{4pth}
\end{align}
as well as
\begin{align}
 c_1^3 &\equiv  \frac{1}{6} \int_{1234} P_{12}^3 \co \ \ \ \ \, c_5^3 \equiv  \frac{1}{2} \int_{1234} P_{12}^2 P_{34} \ , & c_9^3 \equiv    \int_{1234} P_{12} P_{13} P_{23} \phantom{\ .}\notag \\
  c_2^3 &\equiv  \frac{1}{6} \int_{1234} P_{13}^3 \co \ \ \ \ \, c_6^3 \equiv  \frac{1}{2} \int_{1234} P_{12}^2 P_{13} \ , & c_{10}^3 \equiv    \int_{1234} P_{12} P_{13} P_{14}  \phantom{\ .} \notag\\
   c_3^3 &\equiv  \frac{1}{2} \int_{1234} P_{12}^2 P_{23} \co c_7^3 \equiv  \frac{1}{2} \int_{1234} P_{12} P_{13}^2 \ ,& c_{11}^3 \equiv    \int_{1234} P_{12} P_{34} P_{13} \phantom{\ .}\notag \\
   c_4^3 &\equiv  \frac{1}{2} \int_{1234} P_{13}^2 P_{24} \co c_8^3 \equiv  \int_{1234} P_{12} P_{23} P_{34} \ , & c_{12}^3 \equiv    \int_{1234} P_{13} P_{24} P_{12}  \ .  
\label{4pti}
\end{align}
As will be demonstrated in \secref{sec:stringemzv}, eMZVs defined in \eqn{l2.7}
are the natural language to describe the above $c_i^j$ and to understand the
linear combinations appearing after applying momentum conservation \eqn{4ptc}:
\begin{align}
 &I_{4\te{pt}}(1,2,3,4) = c_0 \, + \, 2(c_1^1-c_2^1) \, (s_{12}+s_{23})  \, + \,  (2 c_1^2 + 2 c_2^2 - c_3^2 - c_4^2) \, \big( s_{12}^2 + \tfrac{1}{4}s_{12} s_{23} + s_{23}^2 \big) \notag \\
& \ + \, \frac{1}{4}\, (-2 c_1^2 + 14 c_2^2 + c_3^2 - 7 c_4^2) \, s_{12} s_{23} \, + \, 
 2 \, (c_{10}^3 - 2 c_1^3 - c_2^3 + 2 c_3^3 + c_4^3 - 2 c_9^3) \, s_{12}s_{23} (s_{12}+s_{23})
 \notag \\
 & \  + \,
 ( 2 c_{10}^3 + 2 c_1^3 - 2 c_2^3 + 6 c_3^3 + 2 c_4^3 - 8 c_6^3 - 2 c_8^3 ) \, (s_{12}+s_{23})(s_{12}^2 +s_{12}s_{23}+s_{23}^2) \, + \, {\cal O}(\ap^4)\,.
 \label{4ptj}
\end{align}
A first flavor of relations among $c_i^j$ (and thus ultimately among eMZVs)
can be obtained by exploiting cyclic and reflection properties of five-point
integrals such as 
\begin{equation}
\int_{12345} P_{45} \pa_2 P_{23} = \int_{12345} P_{51} \pa_2 P_{23}  \ \
\ \Rightarrow \ \ \ \int_{1345} P_{45}  P_{13} = \int_{1345} P_{51}  P_{13}
\ \ \ \Rightarrow \ \ \ c_{3}^2 = c_5^2 \ ,
\label{pl30}
\end{equation}
see \eqn{4pte} for the measure $\int_{12345}$. Similar methods imply that
\begin{equation}
2 c_6^2 = c_3^2 + c_4^2 \ , \ \ \ \
c_{3}^3 = c_5^3\ , \ \ \ \
c_{10}^3 = c_{11}^3\ , \ \ \ \
c_{7}^3+c_6^3 = c_3^3+ c_4^3\ , \ \ \ \
c_{11}^3+c_{10}^3 = c_8^3+c_{12}^3 \,,
\label{pl31}
\end{equation}
these relations have been used to eliminate $c_5^2, c_6^2$ as well as
$c_5^3,c_7^3,c_{12}^3, c_{11}^3$ from \eqn{4ptj}.

Note that the $\ap$-expansion of closed string one-loop amplitudes has been
analyzed along similar lines in refs.~\cite{Green:1999pv,Green:2008uj, Richards:2008jg,
Green:2013bza}. Since each closed-string insertion point $z_j$ is integrated
over the entire torus $E_\tau$, integrals involving propagators with a free
endpoint vanish and therefore much fewer closed-string counterparts of the
coefficients $c_i^j$ arise.

\subsection{Elliptic multiple zeta values}
\label{sec:stringemzv}

In this section we convert the constituents of the $\ap$-expansion, $c_i^j$
defined by eqns.~(\ref{4ptg}), (\ref{4pth}) and (\ref{4pti}), to eMZVs.
This will provide a characterization of the particular linear combinations of
$c_{i}^j$ which appear in \eqn{4ptj} along with various powers of $s_{12}$ and
$s_{23}$.

The leading term $c_0$ in \eqn{4ptg} can be straightforwardly
evaluated to yield $\frac{1}{6}$ and furnishes a special case of
\begin{equation}
 \omega(\underbrace{0,0,\ldots,0}_{n})=\frac{1}{n!}    \, ,
\label{cons0}
\end{equation}
which follows from multiple insertions of $1=f^{(0)}(z_i)$. Nevertheless, it
will prove instructive for the comparison with higher orders in $\ap$ to
express $c_0$ as an unevaluated eMZV:
\begin{align}
c_0 &= \int^{1}_0 f^{(0)}(z_4)\, \dd z_4  \int^{z_4}_0 f^{(0)}(z_3)\, \dd z_3  \int^{z_3}_0 f^{(0)}(z_2)\, \dd z_2 \notag \\
&= \GL(0,0,0;1)  = \omega(0,0,0)\,.
\label{4ptk}
\end{align}
Below, we will repeatedly apply the definitions \eqn{eqn:defGell} and
\eqn{l2.7} of \IIEC s and eMZVs, respectively, in order to
express the other integrals $c_i^j$ in the same fashion.

\subsubsection{First order in \texorpdfstring{$\boldsymbol{P_{ij}}$}{Pij}: integrals \texorpdfstring{$\boldsymbol{c_i^1}$}{c1}}

At linear order in $s_{ij}$, we substitute $P_{1j}=\int^{z_j}_0 f^{(1)}(w) \,
\dd w$ according to \eqn{4ptda} and $z_1=0$ into the definitions \eqn{4ptg}
and find
\begin{align}
c_1^1 &=  \int^{1}_0 f^{(0)}(z_4)\, \dd z_4  \int^{z_4}_0 f^{(0)}(z_3)\, \dd z_3  \int^{z_3}_0 f^{(0)}(z_2)\, \dd z_2  \int^{z_2}_0 f^{(1)}(w)\, \dd w \notag \\
&= \GL(0,0,0,1;1) = \omega(1,0,0,0)
\label{4ptl}
\\
c_2^1 &=  \int^{1}_0 f^{(0)}(z_4)\, \dd z_4  \int^{z_4}_0 f^{(0)}(z_3)\, \dd z_3  \int^{z_3}_0 f^{(0)}(z_2)\, \dd z_2  \int^{z_3}_0 f^{(1)}(w)\, \dd w \notag \\
&= \int^{1}_0 f^{(0)}(z_4)\, \dd z_4  \int^{z_4}_0 f^{(0)}(z_3)\, \dd z_3  \, \GL(0;z_3) \, \GL(1;z_3) \notag \\
&= \int^{1}_0 f^{(0)}(z_4)\, \dd z_4  \int^{z_4}_0 f^{(0)}(z_3)\, \dd z_3 \, \big[ \GL(1,0;z_3) + \GL(0,1;z_3) \big]\notag \\
&= \GL(0,0,0,1;1) +\GL(0,0,1,0;1) = \omega(1,0,0,0)+\omega(0,1,0,0) \ .
\label{4ptm}
\end{align}
The second line of \eqn{4ptm} makes use of the shuffle product \eqn{gl2} for
\IIEC s. Equivalence of \eqn{4ptl} with the cyclically shifted integrand
\begin{align}
 \int_{1234} P_{14} &=  \int^{1}_0 f^{(0)}(z_4)\, \dd z_4  \int^{z_4}_0 f^{(0)}(z_3)\, \dd z_3  \int^{z_3}_0 f^{(0)}(z_2)\, \dd z_2  \int^{z_4}_0 f^{(1)}(w)\, \dd w \notag \\
&=\omega(1,0,0,0) + \omega(0,1,0,0)+\omega(0,0,1,0)
\label{4ptn}
\end{align}
can be checked using antisymmetry $\omega(0,1,0,0)+\omega(0,0,1,0)=0$ following
from \eqn{l2.7r}.

\subsubsection{Second order in \texorpdfstring{$\boldsymbol{P_{ij}}$}{Pij}: integrals \texorpdfstring{$\boldsymbol{c_i^2}$}{c2}}

At quadratic order in $s_{ij}$, the rewriting $P_{1j}=\int^{z_j}_0 f^{(1)}(w)
\, \dd w=-\int_{z_j}^1 f^{(1)}(w) \, \dd w$ allows to straightforwardly address
any quadratic monomial in
$P_{12},P_{13},P_{14}$ along the lines of \eqns{4ptl}{4ptm}:
\begin{subequations}
\begin{align}
c_1^2 &= \omega(1,1,0,0,0)  \label{4pto}
\\
c_2^2 &= \omega(1,1,0,0,0)+\omega(1,0,1,0,0)+\omega(0,1,1,0,0) \label{4ptp}
\\
c_3^2 &= -\omega(1,0,0,0,1)  \label{4ptq}
\\
c_6^2 &= 2\omega(1,1,0,0,0)+\omega(1,0,1,0,0) \ .\label{4ptr}
\end{align}
\end{subequations}
Then, \eqns{pl30}{pl31} can be used to determine the remaining two
$c_j^2$ in \eqn{4pth}:
\begin{subequations}
\begin{align}
c_4^2 &=  2\omega(1,1,0,0,0)+\omega(1,0,1,0,0) - \omega(1,0,0,1,0)  \label{4pts}
\\
c_5^2 &= -\omega(1,0,0,0,1)  \ .\label{4ptt}
\end{align}
\end{subequations}
Note that the integration limits $\int^{z_j}_0\!\ldots$ in the representation of $P_{1j}$ can be traded for $-\int_{z_j}^1 \! \ldots$. This is equivalent to applying a shuffle relation \eqn{l2.7sh},
\begin{align}
0 = \omega(1) \omega(1,0,0,0) &= 2\omega(1,1,0,0,0) +\omega(1,0,1,0,0) +\omega(1,0,0,1,0) +\omega(1,0,0,0,1)  
\label{4ptu} 
\\
0 = \omega(1) \omega(0,1,0,0) &= \omega(1,0,1,0,0) +2\omega(0,1,1,0,0) +\omega(0,1,0,1,0) +\omega(0,1,0,0,1)  \ ,
\label{4ptv} 
\end{align}
where $\omega(1)$ vanishes by the reflection identity \eqn{l2.7r}.

\subsubsection{Integration techniques for \texorpdfstring{$\boldsymbol{P_{23},P_{24},P_{34}}$}{P232434}}

Green functions $P_{ij}$ where both indices describe a leg to be integrated 
(legs $2, 3, 4$) are more difficult to integrate. Their integral representation
\eqn{4ptda} inevitably gives rise to iterated integrals $\GLargz{n_1
&\ldots &n_r}{a_1 &\ldots &a_r}$ with the argument appearing in the labels,
that is $a_i=z$.  Integration over $z_3$ and $z_4$ then requires the techniques
of \subsecref{ssec:remove}, in particular the recursion formul\ae{}
\eqn{extra21} to \eqn{extra23}.

The simple corollary $\GLargz{1 &0}{z &0}=-\GLargz{0 &1}{0&0}$ of the
reflection identity \eqn{eqn:reflection} is sufficient to integrate $P_{23}$
and to reproduce \eqn{4ptl} from a different cyclic representative. The
quadratic case $c_5^2= \int_{1234}P_{12} P_{34}$, on the other hand, requires
more effort. One obtains
\begin{align}
\int_{1234} P_{12} P_{34}  &= - \int^1_0 f^{(0)} \dd z_4 \int^{z_4}_0 f^{(1)}(w-z_4) \dd w \int^{w}_0 f^{(0)} \dd z_3 \int^{z_3}_0 f^{(0)} \dd z_2 \int^{z_2}_0 f^{(1)}(u) \dd u \notag \\
&= - \int^1_0 f^{(0)}(z_4) \dd z_4  \, \GLarg{1 &0 &0 &1}{z_4 &0 &0 &0}{z_4}\nnl
&= 2 \omega(1,1,0,0,0) - 2 \omega(2,0,0,0,0) - \omega(0,2,0,0,0) - \zeta_2 \omega(0,0,0) \ ,
\label{4ptva}
\end{align}
where $\GLarg{1 &0 &0 &1}{z_4 &0 &0 &0}{z_4}$ has been reexpressed via
\eqn{lgth5a} in the last step. In order to reproduce the result of \eqn{4ptt},
$-\omega(1,0,0,0,1)$, one needs to combine the shuffle relations
\eqns{4ptu}{4ptv} with \eqns{l2.8x1}{l2.8x2}. The desired result then follows
from the constant eMZVs in \eqn{cons0} and $\omega(2)=-2\zeta_2$ which is a special case of
\begin{equation}
\omega(n)= \left\{ \begin{array}{cc} -2\zeta_n &: \ n \ \te{even} \\ 0 &: \ n \ \te{odd} \end{array} \right.  \ .
\label{cons1}
\end{equation}
The expression for $\omega(n)$ can be inferred from order $q^0$ in the expansions
\eqns{kron7.10}{kron7.11}.

\subsubsection{Third order in \texorpdfstring{$\boldsymbol{P_{ij}}}{Pij}$: integrals \texorpdfstring{$\boldsymbol{c_i^3}$}{c3}}

Starting from the third order in Mandelstam variables, relations such as
\eqn{pl31} are no longer sufficient to reduce the complete list of $c_i^3$ in
\eqn{4pti} to elementary integrals over monomials in $P_{12}$, $P_{13}$ and
$P_{14}$.  Instead, the inevitable factors of $P_{23},P_{24}$ and $P_{34}$ require the 
procedure described in \eqn{4ptva} together with the recursive identities \eqn{extra21} 
to \eqref{extra23} in order to rearrange the labels of the \IIEC s. This allows 
to reduce integrals over arbitrary monomials in $P_{ij}$ with $1\leq i<j\leq 4$ to eMZVs. 
The integrals $c_i^3$, which are cubic in $P_{ij}$, give rise to
\begin{subequations}
\begin{align}
c_1^3 &= \omega(1,1,1,0,0,0) \label{com7}  \\
c_2^3 &= \omega(1,1,1,0,0,0)+\omega(1,1,0,1,0,0) +\omega(1,0,1,1,0,0)+\omega(0,1,1,1,0,0) \label{com8} \\ 
c_3^3 &= -\omega(1,1,0,0,0,1) \label{com9}  \\ 
c_4^3 &= 6\omega(1,1,1,0,0,0)+3\omega(1,1,0,1,0,0) +\omega(1,0,1,1,0,0)+\omega(1,1,0,0,0,1) \label{com10}  \\ 
c_5^3 &= -\omega(1,1,0,0,0,1) \label{com11}  \\ 
c_6^3 &= 3\omega(1,1,1,0,0,0)+\omega(1,1,0,1,0,0)  \label{com12}  \\ 
c_7^3 &= 3\omega(1,1,1,0,0,0)+2\omega(1,1,0,1,0,0) +\omega(1,0,1,1,0,0)
\label{com13}  \\ 
c_8^3 &=2 \omega(2,0,0,0,0,1)+\omega(0,2,0,0,0,1) -2\omega(1,1,0,0,0,1) -\zeta_2 \omega(1,0,0,0) \label{com14}  \\ 
%%%
c_9^3 &= 2 \omega(2,0,0,0,1,0) + 2 \omega(2,0,0,0,0,1) + 
 \omega(0,2,0,0,1,0) +  \omega(0,2,0,0,0,1)  \notag  \\
 %%%
 &\ \ \ \ - 2 \omega(1,1,0,0,1,0) -2 \omega(1,1,0,0,0,1)  -  \zeta_2 \omega(1,0,0,0) -  \zeta_2 \omega(0,1,0,0) \label{com15} \\
c_{10}^3 &= -2\omega(1,1,0,0,0,1)-\omega(1,0,1,0,0,1) \label{com16}  \\ 
 c_{11}^3 &= -2\omega(1,1,0,0,0,1) - \omega(1,0,1,0,0,1) \label{com17}  \\
c_{12}^3 &=-2\omega(2,0,0,0,0,1)- \omega(0,2,0,0,0,1) + \zeta_2 \omega(1,0,0,0) \notag \\
& \ \ \ \ - 2 \omega(1,0,1,0,0,1)-2 \omega(1,1,0,0,0,1)  \ ,\label{com18} 
\end{align}
\end{subequations}
where the occurrences of $\zeta_2$ can be traced back to \eqn{lgth4}.

\subsubsection{Assembling the results}

Momentum conservation only admits particular linear combinations of $c_i^j$ in
the four-point amplitude \eqn{4ptj}. It turns out that for all cases considered divergent eMZVs with the
singular integrand $f^{(1)}$ in the first or last position drop out. Up to third order in $s_{ij}$, we have
\begin{align}
 &I_{4\te{pt}}(1,2,3,4) = \omega(0,0,0) \, - \, 2\omega(0,1,0,0) \, (s_{12}+s_{23})  \, + \,  2 \omega(0,1,1,0,0)\,  \big( s_{12}^2  + s_{23}^2 \big)   \label{4ptja} \\
&\, - \, 2 \omega(0,1,0,1,0) \,s_{12}s_{23} \, + \, \beta_5 \, (s_{12}^3+2 s_{12}^2 s_{23} + 2s_{12} s_{23}^2+s_{23}^3)
\, + \, \beta_{2,3} \, s_{12} s_{23}(s_{12}+s_{23}) \, + \, {\cal O}(\ap^4)
\notag
\end{align}
with 
\begin{align}
\beta_5 &=  \frac{4}{3} \, \big[ \omega(0,0,1,0,0,2)+\omega(0,1,1,0,1,0) - \omega(2,0,1,0,0,0) - \zeta_2 \omega(0,1,0,0) \big]  \\
\beta_{2,3} &= - \frac{1}{3} \omega(0, 0, 1, 0, 2, 0)+ 
 \frac{3}{2} \omega(0, 1, 0, 0, 0, 2) +
 \frac{1}{2} \omega(0, 1, 1, 1, 0, 0) \notag \\
 & \ \ \ +
 2 \omega(2, 0, 1, 0, 0, 0) + 
 \frac{4}{3} \omega(0, 0, 1, 0, 0, 2) +
 \frac{10}{3} \zeta_2 \omega(0, 1, 0, 0) \ ,
\end{align}
and the pattern at higher orders is under investigation. The above
expressions for $\beta_5$ and $\beta_{2,3}$ are obtained using various eMZV
relations using the methods of \subsecref{sec:eMZVrel}.

\subsection{On the \texorpdfstring{$\boldsymbol{q}$}{q}-expansion of eMZVs and the string amplitude}

The evaluation of eMZVs as initiated in \eqn{cons0} and \eqn{cons1} will be
pursued systematically in \cite{WIP,Matthes:edzv,Matthes:thesis}. In this section, we give a
glimpse of non-trivial $q$-dependence in simple cases and provide consistency
checks for the constant piece of the low energy expansion \eqn{4ptja} of the
four-point amplitude.

\subsubsection{The simplest \texorpdfstring{$\boldsymbol{q}$}{q}-expansions}
\label{sec:qexp}

To determine the $q$-expansions of the simplest eMZVs, we start from the
expansions of $f^{(1)}$ and $f^{(2)}$ spelled out in \eqn{kron7.14}, which in
turn is based on \eqns{kron7.6}{kron7.6a}. Using the integrals in
\appref{app:trig}, we arrive at
\begin{equation}
\omega(0,1,0,0) = \frac{\zeta_3}{8\zeta_2} + \frac{3}{2\pi^2} \sum_{m,n=1}^{\infty} \frac{1}{m^3} q^{mn}  
\label{explq1}
\end{equation}
as well as
\begin{align}
\omega(0,1,1,0,0) &= \frac{\zeta_2}{15}-\frac{1}{2\pi^2} \sum_{m,n=1}^{\infty} \frac{n}{m^4} q^{mn}+ \frac{1}{3} \sum_{m,n=1}^{\infty} \frac{n}{m^2} q^{mn}
\label{explq2} 
\\
%%%
\omega(0,1,0,1,0) &= -\frac{\zeta_2}{60}+\frac{2}{\pi^2} \sum_{m,n=1}^{\infty} \frac{n}{m^4} q^{mn}- \frac{1}{3} \sum_{m,n=1}^{\infty} \frac{n}{m^2} q^{mn} \ .
\label{explq3}
\end{align}
A systematic method is under investigation and will appear in
\cite{Matthes:thesis}. Note that the $q$-dependence of all the examples above
can be expressed in terms of the function $\ELi_{n,m}$ introduced in section~8
of ref.~\cite{Adams:2014vja} at arguments $x=y=1$.

\subsubsection{The constant piece of eMZVs and the \texorpdfstring{$\boldsymbol{\ap}$}{ap}-derivative}

The $t$-integration in the four-point amplitude \eqn{4pta} is divergent unless
the choice of gauge group $SO(32)$ leads to cancellations between the cylinder
and the M\"obius-strip diagram \cite{Green:1984ed}. The divergence is
interpreted as a zero-momentum dilaton propagating to the vacuum and therefore
proportional to the derivative of the tree level amplitude with respect to $\ap$
\cite{Green:1981ya}. The latter is given by
\begin{equation}
A_{\te{string}}^{\te{tree}}(1,2,3,4) =
\frac{\Gamma(1+s_{12}) \Gamma(1+s_{23}) }{\Gamma(1+s_{12}+s_{23})} A_{\te{YM}}^{\te{tree}}(1,2,3,4)
\label{cons2a} 
\end{equation}
with $\ap$-expansion  
\begin{align}
&\frac{\Gamma(1+s_{12}) \Gamma(1+s_{23}) }{\Gamma(1+s_{12}+s_{23})} =
\exp \left\{ \sum_{k=2}^{\infty} (-1)^k \frac{ \zeta_k }{k}  \big[ s_{12}^k+s_{23}^k - (s_{12}+s_{23})^k \big] \right\}
\notag \\
& \ \ = 1 - \zeta_2 s_{12} s_{23} + \zeta_3 s_{12} s_{23}(s_{12}+s_{23}) - \zeta_4 s_{12} s_{23} \big(s_{12}^2+\tfrac{1}{4} s_{12} s_{23} + s_{23}^2 \big) \label{cons2b}  \\
& \ \ \ \ \ \ \ \, +\zeta_5 s_{12} s_{23} (s_{12}^3 + 2 s_{12}^2 s_{23}+2 s_{12}s_{23}^2+s_{23}^3) -\zeta_2\zeta_3 (s_{12} s_{23})^2 (s_{12}+s_{23}) + {\cal O}(\ap^6) \ .
\notag
\end{align}
In the representation of the one-loop amplitude given in \eqn{4ptja}, the
divergence originates from the constant part of the eMZVs' power series expansion
in $q=e^{2\pi i \tau}=e^{-2\pi t}$. A systematic method to extract the
constant term of eMZVs will be described in ref.~\cite{WIP}. The resulting
divergence in the above result is given by
\begin{align}
&A_{\te{string}}^{\te{1-loop}}(1,2,3,4) \, \Big|_{\te{div}} =  s_{12}s_{23} A_{\te{YM}}^{\te{tree}}(1,2,3,4) \, I_{4\te{pt}}(1,2,3,4) \, \Big|_{q^0} \notag \\
& \ \ =  \frac{1}{2\pi^2} \ s_{12}s_{23} A_{\te{YM}}^{\te{tree}}(1,2,3,4) \Big\{ 2 \zeta_2 - 3 \zeta_3 (s_{12}+s_{23}) + 4 \zeta_4 \big(s_{12}^2+\tfrac{1}{4} s_{12} s_{23} + s_{23}^2 \big) \label{cons3} \\
& \ \ \ \ \ \  - 5 \zeta_5 (s_{12}^3 + 2 s_{12}^2 s_{23}+2 s_{12}s_{23}^2+s_{23}^3) +5 \zeta_2\zeta_3 s_{12} s_{23} (s_{12}+s_{23}) + {\cal O}(\ap^4) 
\Big\} \ , \notag
\end{align}
which is consistent with the $\ap$-derivative of the tree amplitude
\cite{Green:1981ya} upon comparison with \eqn{cons2b},
\begin{equation}
A_{\te{string}}^{\te{1-loop}}(1,2,3,4) \, \Big|_{\te{div}} =  -{\ap \over 2\pi^2} \frac{\partial}{\partial \ap} A_{\te{string}}^{\te{tree}}(1,2,3,4)  \ .
 \label{cons4}
\end{equation}

%%%%%%%%%%%%%%%%%%%%%%%%%%%%%%%%%%%%
%%%%%%%%%%%%%%%%%%%%%%%%%%%%%%%%%%%%
%%%%%%%%%%%%%%%%%%%%%%%%%%%%%%%%%%%%
%%%%%%%%%%%%%%%%%%%%%%%%%%%%%%%%%%%%
%%%%%%%%%%%%%%%%%%%%%%%%%%%%%%%%%%%%
%%%%%%%%%%%%%%%%%%%%%%%%%%%%%%%%%%%%

\section{Multi-particle one-loop string amplitudes and \texorpdfstring{$\boldsymbol{f^{(n)}}$}{fn}}
\label{sec:multipt}

This section is devoted to one-loop amplitudes involving five and more open
string states. We firstly provide the five-point extension of the four-point
$\ap$-expansion in \eqn{4ptja}. It is secondly demonstrated that the doubly-periodic
functions $f^{(n)}$ defined in \secref{sec:math} naturally enter the
calculation of one-loop amplitudes with any number of external legs.

\subsection{The five-point open string amplitude}

In the same way as the four-point open string amplitude in \eqn{4pta} allows to
factor out the polarization dependence via $A_{\te{YM}}^{\te{tree}}(1,2,3,4)$,
one can express the five-point string amplitude in a basis of color-ordered
trees of YM theory \cite{Mafra:2012kh}. BCJ relations \cite{Bern:2008qj} single
out two independent subamplitudes $A_{\te{YM}}^{\te{tree}}(1,\rho(2,3),4,5)$
with permutation $\rho \in S_2$, and for convenience, we consider the same
color orderings in the one-loop string theory counterparts:
\begin{equation}
A_{\te{string}}^{\te{1-loop}}(1,\sigma(2,3),4,5) =   \int_0^{\infty} \dd t \ \sum_{\rho \in S_2} I_{5\te{pt}}(\sigma|\rho) A_{\te{YM}}^{\te{tree}}(1,\rho(2,3),4,5)  \,.
\label{5pta} 
\end{equation}
The $2\times 2$ matrix $I_{5\te{pt}}(\sigma|\rho)$ is the generalization of
the four-point scalar integral $I_{4\te{pt}}(1,2,3,4)$. It can be assembled
from the kinematic factors which were simplified in
ref.~\cite{Mafra:2012kh} using the pure spinor formalism \cite{Berkovits:2000fe},
\begin{align}
\sum_{\rho \in S_2} I_{5\te{pt}}(1|\rho) A_{\te{YM}}^{\te{tree}}(1,\rho(2,3),4,5) &= \int_{12345} \prod_{k<l}^5 \exp\big[ s_{kl} P_{kl} \big] 
\label{5ptb} \\ 
& \ \ \ \ \ \ \  \times \big[ s_{23} f^{(1)}_{23} \langle C_{1|23,4,5} \rangle + (23 \leftrightarrow 24,25,34,35,45) \big]  \notag \\
%%%
 \langle C_{1|23,4,5} \rangle &= s_{45} \, \bigl(s_{24} \,A_{\te{YM}}^{\te{tree}}(1,3, 2, 4, 5)  -  s_{34}
   A_{\te{YM}}^{\te{tree}}(1,2, 3, 4, 5) \bigr) \ .
\label{5ptc} 
\end{align}
The integration measure $\int_{12345}$ is defined in \eqn{4pte}, the functions
$f^{(1)}_{ij} =f^{(1)}(z_{i}-z_{j})$ stem from OPE contractions among the
worldsheet fields and the five-point Mandelstam invariants \eqn{4ptb} can be
cast into a five-dimensional basis via momentum conservation, e.g.
$s_{13}=s_{45}-s_{12}-s_{23}$.

From the mathematical point of view, the only novel five-point ingredient as
compared to the four-point amplitude is the extra factor of
$f^{(1)}_{ij}=\partial P_{ij}$ in the integrand of \eqn{5ptb}. Thanks to the
embedding of $f^{(1)}$ into the framework of \IIEC s \eqn{eqn:defGell}, the
$\ap$-expansion of the integrals $\int_{12345} f^{(1)}_{ij} \prod_{k<l}^5
\exp\big[ s_{kl} P_{kl} \big] $ in \eqn{5ptc} is again captured by eMZVs. The
detailed discussion of kinematic poles as well as the order-by-order treatment
of the exponential will be discussed elsewhere; here we simply quote the final
result:
\begin{align}
I_{5\te{pt}}(\sigma|\rho) &= \big[ - \omega(0,0,0) P_2 -2 \omega(0,1,0,0)M_3  - 5 \omega(0,1,1,0,0) P_4 \nnl
& \ \ \ \ \ \ \ \ - \big(2 \omega(0,1,0,1,0) + \tfrac{1}{2} \omega(0,1,1,0,0) \big) L_4 +  {\cal O}(\alpha'^5)\big]_{\sigma,\rho}\ .
\label{5ptd}
\end{align}
Up to weight two at order ${\cal O}(\alpha^4)$, the eMZV content is the same as
in the four-point expansion \eqn{4ptja}. The accompanying $2\times 2$ matrices
$P_i,M_i,L_i$ are indexed by permutations $\rho, \sigma$, and their entries are
polynomials of degree $i$ in Mandelstam variables. The representatives $P_i$
and $M_i$ already appear in the $\ap$-expansion of open-string tree amplitudes,
along with even and odd Riemann zeta values $\zeta_i$, respectively
\cite{Schlotterer:2012ny}. Given that the low-energy limit of
one-loop amplitudes at any multiplicity has the mass dimension of $s_{ij}^2
A_{\te{YM}}^{\te{tree}}(\ldots)$, the eMZV coefficients of $P_i,M_i,L_i$ have
weight $i-2$. This amounts to a shift of $-2$ in weight in comparison to the
MZV coefficients of $P_i,M_i$ at tree level.

They are available at the website \cite{MZVWebsite}
whereas $L_4$ reads
\begin{align}  
(L_4)_{11} &= s_{12}^2 s_{23}^2 + 2 s_{12}^2 s_{23} s_{24} + s_{12}^2 s_{24}^2 + 2 s_{12}^2 s_{23} s_{34 }+ 
 2 s_{12} s_{13} s_{23} s_{34} + 2 s_{12} s_{23}^2 s_{34} \notag\\
 & \ \ \ \ \ \ \ \ \ \ + 2 s_{12}^2 s_{24} s_{34} + 
 s_{12} s_{13} s_{24} s_{34} + 2 s_{12} s_{23} s_{24} s_{34} + s_{12}^2 s_{34}^2 + 2 s_{12}s_{13} s_{34}^2\notag \\
 & \ \ \ \ \ \ \ \ \ \  +
  s_{13}^2 s_{34}^2 + 2 s_{12} s_{23} s_{34}^2 + 2 s_{13} s_{23} s_{34}^2 + s_{23}^2 s_{34}^2  \label{pl34} \\
(L_4)_{12} &=  -s_{13} s_{24} (3 s_{12} s_{23} + s_{13} s_{23} + s_{23}^2 + 2 s_{12} s_{24} + s_{13} s_{24} + 
   s_{23} s_{24} \notag \\
 & \ \ \ \ \ \ \ \ \ \   + 3 s_{12} s_{34} + 2 s_{13} s_{34} + 3 s_{23} s_{34}) 
\end{align}
and $(L_4)_{22}=(L_4)_{11}  \big|_{2\leftrightarrow 3}$ and
$(L_4)_{21}=(L_4)_{12}  \big|_{2\leftrightarrow 3}$. The relabelling
$2\leftrightarrow 3$ refers to the $i,j$ along with the Mandelstam invariants
$s_{ij}$.

The four-point one-loop amplitude \eqn{4ptja} can be cast into the same form as
\eqn{5ptd} upon setting $L_4 \rightarrow 0$ and
\begin{align}
P_2 & \rightarrow  - s_{12}s_{23} \, , \ \ \ \ M_3  \rightarrow  s_{12}s_{23} \,(s_{12}+s_{23}) \, , \ \ \ \ P_4  \rightarrow  -\frac{2}{5}\,s_{12}s_{23} \, \left(s_{12}^2+\tfrac{1}{4}s_{12}s_{23}+s_{23}^2 \right) \ , \label{pl14}
% \\
%M_5 & \rightarrow  su(s^3+2s^2u+2su^2+u^3) = su(s+u)(s^2+su+u^2)  \co L_i \rightarrow 0\ ,  \notag
\end{align}
in agreement with the four-point open string tree \eqn{cons2a}. The pattern of
eMZVs at higher orders in $\ap$ as well as the properties of the novel matrices
$L_i$ are left for further projects.

\subsection{Functions \texorpdfstring{$\boldsymbol{\f{n}}$}{fn} from the RNS formalism}

In this subsection we will show that the doubly-periodic functions $\f{n}$ for any $n$
are naturally generated in the one-loop amplitude computation using the RNS
formalism \cite{Ramond:1971gb, Neveu:1971rx, Neveu:1971iv}. Their emergence in
the parity-even and parity-odd sectors turns out to follow two separate
mechanisms.

\subsubsection{Parity-even RNS amplitudes}

In the parity-even sector of the RNS computation, the functions $f^{(n)}$ arise
from the summation over the even spin structures of the fermions on a genus-one
worldsheet.  We also take this opportunity to use the method of
refs.~\cite{Tsuchiya:1988va,Tsuchiya:2012nf} to write down explicit results for
the $\nol$-point spin sum for $\nol>7$.

\paragraph{Definition of \texorpdfstring{$\boldsymbol{V_p(x_1, \ldots,x_\nol)}$}{Vn}.} In the
subsequent we use the variables $x_i\equiv z_i - z_{i+1}$ for $i=1,
\ldots,\nol$ with the condition $z_{\nol+1}=z_1$ such that $\sum_{i=1}^\nol x_i
=0$.  Using the shorthand $\Omega_i\equiv \a\Omega(x_i,\alpha)$ it follows from
\eqn{alt15} that the $\a^{p}$-component of $\Omega_1 \cdots \Omega_\nol$ has
at most $p$ simultaneous single poles in the variables
$x_i$.  This suggests the following definition
\begin{equation}
\label{Vpdef}
V_{p}(x_1,x_2, \ldots,x_\nol) \equiv (\Omega_1\Omega_2 \ldots\Omega_\nol)\Big|_{\a^p}\,.
\end{equation}
For example, with $f_i^{(n)} \equiv f^{(n)}(x_i)$,
\begin{align}
V_1(x_1,\ldots,x_5) &= \sum_{i=1}^5 f_i^{(1)} \notag \\
 V_2(x_1,\ldots,x_6) &= \sum_{i=1}^6 f_i^{(2)} + \sum_{1\le i< j}^6f^{(1)}_i f^{(1)}_j\notag \\
V_3(x_1,\ldots,x_7) &=  \sum_{i=1}^7  f_i^{(3)} + \sum_{1\leq i<j}^7 ( f_i^{(1)} f_j^{(2)}+ f_i^{(2)} f_j^{(1)} )
+ \sum_{1\leq i<j<k}^7f_i^{(1)} f_j^{(1)} f_k^{(1)}\notag\\
V_4(x_1,\ldots,x_8) &=  \sum_{i=1}^8 f^{(4)}_i + \sum_{1\leq i<j}^8 (f^{(1)}_i f^{(3)}_j +f^{(2)}_i f^{(2)}_j +f^{(3)}_i f^{(1)}_j ) +  \sum_{1\leq i<j<k<l}^8 f^{(1)}_i f^{(1)}_j f^{(1)}_k f^{(1)}_l\notag \\
&+  \sum_{1\leq i<j<k}^8 (f^{(1)}_i f^{(1)}_j f^{(2)}_k +f^{(1)}_i f^{(2)}_j f^{(1)}_k +f^{(2)}_i f^{(1)}_j f^{(1)}_k ) \ .
\end{align}
Interestingly, the anti-holomorphic recursion \eqn{alt12} implies that
$V_p(x_1, \ldots,x_\nol)$ is holomorphic; ${\p\over\p {\bar z}_i} V_p(x_1,
\ldots,x_N) = 0$. Equivalently, the non-holomorphic factors $\Im(x_i)$ in
$V_{p}(x_1,\ldots,x_\nol)$ trivially vanish because of the condition
$\sum_{i=1}^\nol x_i =0$.  One can therefore replace $\cE_1(x,\tau)$ by
$E_1(x,\tau)$ and $f^{(n)}_i\rightarrow g^{(n)}_i$ in the notation of
\subsecref{sec:gf} to establish manifest holomorphicity. 

Note that the functions in \eqn{Vpdef} were also used in \cite{Dolan:2007eh} to
cast one-loop correlation functions among arbitrary numbers of Kac-Moody currents into a closed form.

\paragraph{Spin sums in one-loop amplitudes.} In the computation of parity-even
one-loop amplitudes in the RNS formalism the bosonic worldsheet fields can be 
straightforwardly integrated out to yield products of $f^{(1)}$, possibly after integration 
by parts. Worldsheet fermions, on the other hand, give rise to the following
spin sums,
\begin{equation}
\label{spinsum}
{\cal G}_\nol(x_1,\ldots,x_\nol) \equiv \sum_{\nu=1,2,3} (-1)^{\nu} 
  \left( { \theta_{\nu+1}(0,\tau) \over \theta'_1(0,\tau) } \right)^4 S_\nu(x_1) S_\nu(x_2) \ldots S_\nu(x_\nol)\,,
\end{equation}
where $\sum_{i=1}^N x_i = 0$, $S_\nu$ is the Szeg\"o kernel and $\nu$ denotes the even spin structure
with associated Jacobi theta functions $\theta_2,\theta_3,\theta_4$ \cite{Namazie:1986zu,igusa, fay, mumford1984tata},
\begin{equation}
\label{szego}
S_\nu(z) \equiv { \theta_1'(0,\tau) \theta_{\nu+1}(z,\tau) \over  \theta_{\nu+1}(0,\tau) \theta_1(z,\tau)}  \ .
\end{equation}
A method to evaluate such sums was presented in ref.~\cite{Tsuchiya:1988va} and
its explicit results at $\nol \le 7$ can be written in terms of $\f1(z)$, the
Weierstrass function $\wp(z)$ and its derivatives $\pd^k\wp(z)$,
\begin{align}
{\cal G}_4(x_1,\ldots,x_4) &=  1
\notag \\
{\cal G}_5(x_1,\ldots,x_5) &=  \sum_{j=1}^5 f^{(1)}(x_j)
\notag \\
{\cal G}_6(x_1,\ldots,x_6) &=  {1\over 2} \bigg\{ \Big( \sum_{j=1}^6 f^{(1)}(x_j) \Big)^2 -
\sum_{j=1}^6 \wp(x_j) \bigg\}
\notag \\
{\cal G}_7(x_1,\ldots,x_7) &=  {1\over 6} \bigg\{ \Big( \sum_{j=1}^7 f^{(1)}(x_j) \Big)^3 - 
\sum_{j=1}^7 \pd \wp(x_j) - 3 \Big( \sum_{j=1}^7 f^{(1)}(x_j) \Big) \Big( \sum_{j=1}^7  \wp(x_j) \Big) \bigg\}   \ .
\label{kron1.5}
%
%{\cal G}_8(x_1,\ldots,x_8) &=   \sum_{(m,n) \neq (0,0)} \frac{const}{(m+n\tau)^4 } + {1\over 24} \bigg\{ \Big( \sum_{j=1}^8 f^{(1)}(x_j) \Big)^4 - \sum_{j=1}^8 \pd^2 \wp(x_j) + 3 \Big( \sum_{j=1}^8  \wp(x_j) \Big)^2 \nnl
%& \ \ \ \ \ - 4 \Big( \sum_{j=1}^8 f^{(1)}(x_j) \Big) \Big( \sum_{j=1}^8 \pd \wp(x_j) \Big) - 6 \Big( \sum_{j=1}^8 f^{(1)}(x_j) \Big)^2 \Big( \sum_{j=1}^8  \wp(x_j) \Big) \bigg\}   
%\label{kron1.7}
\end{align}
One can show that the above results are naturally described by the elliptic
functions $V_p(x_1, \ldots,x_\nol)$,
\begin{equation}
\label{kron1.3}
{\cal G}_\nol(x_1, \ldots,x_\nol) = V_{\nol-4}(x_1, \ldots,x_\nol), \ \ \ \quad 4\le \nol \le 7\,.
\end{equation}
% \begin{align}
% {\cal G}_5(x_1, \ldots,x_5) &= V_1(x_1, \ldots,x_5) = \sum_{i=1}^5 f_i^{(1)}\notag \\
% {\cal G}_6(x_1, \ldots,x_6) &= V_2(x_1, \ldots,x_6) = \sum_{i=1}^6 f_i^{(2)} + \sum_{1\le i< j}f^{(1)}_i f^{(1)}_j \label{kron1.3} \\
% {\cal G}_7(x_1, \ldots,x_7) &= V_3(x_1, \ldots,x_7) =  \sum_{i=1}^7  f_i^{(3)} + \sum_{1\leq i<j}^7 ( f_i^{(1)} f_j^{(2)}+ f_i^{(2)} f_j^{(1)} )
% + \sum_{1\leq i<j<k}^7f_i^{(1)} f_j^{(1)} f_k^{(1)} \ .\notag
% \end{align}
%
An alternative method was used in \cite{Stieberger:2002fh, Stieberger:2002wk} to express ${\cal G}_{N}$ in terms of single derivatives of the bosonic Green function. The equivalence of the expression for ${\cal G}_6$ given in these references with \eqn{kron1.5} can be verified through the Fay identity \eqn{kron1.34}.

Although the results for $\nol\ge 8$ were not written down explicitly in
ref.~\cite{Tsuchiya:1988va}, they also take a natural form when expressed in
terms of elliptic functions $V_p(x_1, \ldots,x_\nol)$,
\begin{align}
\label{higherGs}
{\cal G}_8(x_1, \ldots,x_8) &= V_4(x_1, \ldots,x_8) + 3e_4\notag \\
{\cal G}_9(x_1, \ldots,x_9) &= V_5(x_1, \ldots,x_9) + 3e_4V_1(x_1, \ldots,x_9)\notag \\
{\cal G}_{10}(x_1, \ldots,x_{10}) &= V_6(x_1, \ldots,x_{10}) + 3e_4V_2(x_1, \ldots,x_{10}) + 10e_6\notag\\
{\cal G}_{11}(x_1, \ldots,x_{11}) &= V_7(x_1, \ldots,x_{11}) + 3e_4V_3(x_1, \ldots,x_{11}) + 10e_6 V_1(x_1, \ldots,x_{11})\notag \\
{\cal G}_{12}(x_1, \ldots,x_{12}) &= V_8(x_1, \ldots,x_{12}) + 3e_4V_4(x_1, \ldots,x_{12}) + 10e_6 V_2(x_1,
\ldots,x_{12}) + 42 e_8 \ .
\end{align}
The factors of the Eisenstein series $e_j$ \eqn{eiss} can be systematically computed 
as well. Following ref.~\cite{Tsuchiya:2012nf}, we define  $Q_0(\wp) = 1$, $Q_1(\wp) =
\wp$ and $Q_{k+1}(\wp)=\wp^{(2k)}$. For example,
\begin{align}
\label{wpprimes}
Q_2(\wp) &= 3!\wp^2 - \half g_2 \cr
Q_3(\wp) &= 5!\wp^3 - 18 g_2\wp - 12g_3\cr
Q_4(\wp) &= 7!\wp^4 - 1008 g_2\wp^2 - 720 g_3\wp + 9 g_2^2\cr
Q_5(\wp) &= 9!\wp^5 - 90720g_2\wp^3 - 64800 g_3\wp^2 + 3024 g_2^2 \wp + 2376 g_2 g_3\,,
\end{align}
where the Weierstrass equation $(\wp')^2 = 4\wp^3 - g_2 \wp - g_3$ has been
used to rewrite the $2k^{\rm th}$ derivative of $\wp$ as a polynomial in $\wp$.
In the above equation, $g_2 = - 4(s_1s_2+s_2s_3+s_3s_1)=60e_4$, $g_3 =4s_1
s_2s_3=140e_6$ are the elliptic invariants and $s_i$ are the branch
points of the genus-one elliptic curve $y^2 = 4(z-s_1)(z-s_2)(z-s_3)$
satisfying $s_1+s_2+s_3=0$. Defining
\begin{equation}
\label{Fkdef}
F_{2k-4} \equiv - {1\over (2k-1)!}{\big[ (s_1-s_3)Q_k(s_2) + (s_3-s_2)Q_k(s_1) +
(s_2-s_1)Q_k(s_3)\big] \over (s_1-s_3)(s_3-s_2)(s_2-s_1)},\quad k\ge 4
\end{equation}
straightforward calculation leads to\footnote{The Eisenstein series
$e_8$, $e_{10}$ and $e_{12}$ can be written in terms of $e_4$ and $e_6$ as
follows
$$
e_8 = {3\over 7}e_4^2,\quad e_{10} ={5\over 11}e_4e_6,\quad e_{12}={18\over 143}e_4^3+{25\over 143}e_6^2.
$$
The general formula is written in terms of $d_k\equiv  (2k+3)k! e_{2k+4}$
$$
d_{n+2} = {3n+6\over 2n+9}\sum_{k=0}^n {n\choose k}d_k d_{n-k} \ .
$$
}
$$
F_4 = 3e_4,\quad
F_6 = 10e_6,\quad
F_8 = 42e_8,\quad
F_{10} = 168e_{10},\quad
F_{12} = 627e_{12}+9e_4^3\ ,
$$
which precisely captures the factors of $e_j$ in \eqn{higherGs}. We have
explicitly checked up to $\nol=12$ that the spin sums can be uniformly written as,
\begin{equation}
\label{conj}
{\cal G}_\nol(x_1, \ldots,x_\nol) = V_{\nol-4}(x_1, \ldots,x_\nol) + \sum_{k=1}^{\lfloor{\nol-8\over 2}\rfloor
+1}F_{2k+2}V_{\nol-2k-6}(x_1, \ldots,x_\nol)\,.
\end{equation}

\subsubsection{Parity-odd RNS amplitudes}

The parity-odd sector of the RNS computation entirely stems from the unique odd
spin structure at genus one where the worldsheet spinors obey anti-periodic
boundary conditions along both torus cycles and acquire a zero mode.  The
worldsheet integrand is governed by zero-mode saturation and, probably as a
common feature with the Green-Schwarz or pure spinor formalism, OPE
contractions of the worldsheet fields which generate $N-4$ factors of
$f^{(1)}_{ij}$ where $f^{(n)}_{ij} \equiv f^{(n)}(z_{i}-z_{j})$.

For six points, the direct evaluation of the OPEs gives rise to a quadratic
factor $f^{(1)}_{ij}f^{(1)}_{kl}$ for various combinations of labels capturing
the behavior of the singularities as the vertices collide. However, we know
from the Fay identity \eqn{kron1.34} that these quadratic combinations are not
linearly independent and therefore one is naturally led to higher-weight
$\f{n}$'s when considering a minimal basis of integrals to evaluate. The
simplest example where a higher-weight $\f{n}$ is generated this way is
$\f1_{12}\f1_{13} + \f1_{23}\f1_{21} + \f1_{31}\f1_{32} = \f2_{12} + \f2_{23} +
\f2_{31}$ which can be viewed as generalizing the genus-zero partial fraction
identity \eqn{eqn:partialfraction}. The non-vanishing of the right-hand side
provides an important distinction between one-loop and tree-level string
amplitudes and it is ultimately related to the gauge anomaly cancellation
mechanism in the superstring \cite{Green:1984sg, Green:1984qs}. It can be shown
that the parity-odd part of the six-point amplitude as firstly computed in
ref.~\cite{Clavelli:1986fj} can be entirely written in terms of $\f2$,
i.e.~that any appearance of $f^{(1)}$ can be removed via \eqn{kron1.34}.

More generally, the $\nol-4$ powers of $\f1$ in the $\nol$-point amplitude
allow, via the Fay identity, the generation of $\f{p}$ with up to $p=\nol-4$.
In this way the need for a general integration method for the type of iterated
integrals on an elliptic curve considered in this paper is justified.

%%%%%%%%%%%%%%%%%%%%%%%%%%%%%%%%
%%%%%%%%%%%%%%%%%%%%%%%%%%%%%%%%
%%%%%%%%%%%%%%%%%%%%%%%%%%%%%%%%
%%%%%%%%%%%%%%%%%%%%%%%%%%%%%%%%

\section{Discussion and further directions}

In this article, we have proposed an organization scheme for \IIECl{} and
elliptic multiple zeta values (eMZVs), where the key definitions are provided
in \eqns{eqn:defGell}{l2.7}. The infinite family of doubly-periodic functions
$f^{(n)}$ appearing in the integrands of \secref{sec:polylog} are put into a
mathematical context and are related to multiple elliptic polylogarithms in
\secref{sec:math}. As a first natural and simple application of this framework,
we have identified eMZVs in the $\ap$-expansion of one-loop scattering
amplitudes in open string theory. The leading orders in the low-energy
behavior of the four- and five-point amplitudes in terms of eMZVs are
presented in \eqns{4ptja}{5ptd}. Divergent eMZVs turn out to cancel from our
results.

Having demonstrated the potential of the formalism for an initial example,
there are numerous open questions to be pursued in the near future. Most
obviously, the eMZV content of the low energy expansion of cylinder amplitudes
needs to be understood for higher orders in $\ap$, which can be done
conveniently using the new techniques. Furthermore, the contributions from the
cylinder configuration with open string insertions on both boundaries as well
as from the M\"obius-strip topology shall be determined in terms of 
the iterated integrals introduced in \subsecref{ssec:MEP}.
The $q$-expansion of eMZVs exemplified in \secref{sec:qexp}
offers a promising approach to systematically perform the $t$-integration in
\eqn{4pta} after summing all topologies for the gauge group
$SO(32)$~\cite{Green:1984ed}.

On the mathematical side, the network of relations between eMZVs explored in
\subsecref{sec:eMZVrel} will be further investigated in
refs.~\cite{WIP,Matthes:edzv,Matthes:thesis}. A suitable coaction along the
lines of refs.~\cite{Goncharov:2001iea, Goncharov:2005sla, Brown:2011ik,
Duhr:2012fh, BrownLev} might lead to a natural basis choice for eMZVs and might
allow to further identify patterns in the one-loop string amplitudes. In the
same way as the Drinfeld associator was instrumental in understanding the
pattern of MZVs \cite{Schlotterer:2012ny} in open string tree-level amplitudes
\cite{Drummond:2013vz} and finally allowed to completely determine their
$\ap$-expansion in ref.~\cite{Broedel:2013aza}, the elliptic associators
discussed in ref.~\cite{Enriquez:ellass} might encode the structure of the
$\ap$-expansion at one-loop.  Furthermore, in refs.~\cite{multmod, Brown:mmv}
so-called \emph{multiple modular values} are discussed whose possible relation
to the eMZVs studied here needs to be explored.

In multi-particle one-loop open string amplitudes, the pure spinor formalism, in
particular the ingredients of ref.~\cite{Mafra:2014gsa} are expected to yield a
compact description of the kinematic factors associated to the functions
$f^{(n)}$. While the precise superspace kinematic factors along with various
powers of $f^{(1)}$ have been derived in ref.~\cite{Mafra:2012kh}, the
kinematic companions of $\f{n\geq 2}$ in the higher-point amplitudes are
currently under investigation.

Finally, it would be desirable to find a similar scheme for organizing the
$\ap$-expansion of closed string one-loop amplitudes. In particular, the
worldsheet integrals investigated in refs.~\cite{Green:2008uj, Richards:2008jg,
Green:2013bza} might allow for a description in terms of eMZVs and their
counterpart defined with respect to the other cycle of the torus.  The peculiar
linear combinations of torus integrals appearing in the $\ap$-expansion of
closed-string amplitudes call for an explanation along the lines of the above
finding that divergent eMZVs drop out from the open-string expansions.

%%%%%%%%%%%%%%%%%%%%%%%%%%%%%%%%
%%%%%%%%%%%%%%%%%%%%%%%%%%%%%%%%
%%%%%%%%%%%%%%%%%%%%%%%%%%%%%%%%
%%%%%%%%%%%%%%%%%%%%%%%%%%%%%%%%

\subsection*{Acknowledgments}

OS and NM are grateful to the organizers of the conference ``Numbers and
Physics'' in Madrid in September 2014, in particular to Jos\'e Burgos for
encouraging discussions. We would like to thank Francis Brown, Claude Duhr and
Michael Green for comments and suggestions on the draft of this article as well
as an unknown referee for his elaborate feedback on the first submitted
version.  JB wants to thank the Stanford Institute for Theoretical Physics,
where part of this work was completed, for hospitality.  Furthermore we are
grateful to Francis Brown and Michael Green for helpful discussions and Ulf
K\"uhn for valuable advice.  JB, NM and CRM want to thank the
Albert-Einstein-Institute for hospitality.  OS is grateful to the Department of
Applied Mathematics and Theoretical Physics of the University of Cambridge and
to ETH Z\"urich for hospitality. CRM and OS acknowledge financial support by
the European Research Council Advanced Grant No. 247252 of Michael Green. 

%%%%%%%%%%%%%%%%%%%%%%%%%%%%%%%%
%%%%%%%%%%%%%%%%%%%%%%%%%%%%%%%%
%%%%%%%%%%%%%%%%%%%%%%%%%%%%%%%%
%%%%%%%%%%%%%%%%%%%%%%%%%%%%%%%%

\section*{Appendix}
\appendix

\section{Derivatives of multiple polylogarithms w.r.t.~the labels} 
\label{app:MPrelations}

The proof of the recursion in \eqn{eqn:GIdentity} relies on the derivatives of
multiple polylogarithms \eqn{eqn:defG} with respect to their labels
$a_1,a_2,\ldots,a_n$ \cite{Goncharov:2001iea}:
\begin{align}
  \frac{\pd}{\pd z}G(\vec{a};z)&=\frac{1}{z-a_1}G(a_2,\ldots,a_n;z).
  \label{eqn:derivativeG1}
\\
  \frac{\pd}{\pd a_i}G(\vec{a};z)&=\frac{1}{a_{i-1}-a_i}G(\ldots,\hat{a}_{i-1},\ldots;z) \ 
  + \ \frac{1}{a_i-a_{i+1}}G(\ldots,\hat{a}_{i+1},\ldots;z)\nnl
  & \ \ \ \ \ \ \ \ \ -\frac{a_{i-1}-a_{i+1}}{(a_{i-1}-a_i)(a_i-a_{i+1})}G(\ldots,\hat{a}_i,\ldots;z) \ , \ \ \ \ i \neq 1,n
  \label{eqn:derivativeG}
\\
  \frac{\pd}{\pd a_n}G(\vec{a};z)&=\frac{1}{a_{n-1}-a_n}G(\ldots,\hat{a}_{n-1},a_n;z)
  -\frac{a_{n-1}}{(a_{n-1}-a_n)a_n}G(\ldots,a_{n-1};z) \ .
   \label{eqn:derivativeG2}
\end{align}

\section{Identities for iterated integrals}
\label{ellpolyexpl}

This appendix provides further relations to integrate \IIEC s whose argument
occurs in the labels.

\subsection{Total derivatives}
\label{ellpolyexpla}

The following identities generalize eqns.~\eqref{tzero1} to
\eqref{tzero3} for multiple successive occurrences of the argument $t_0$
in the label. If the first $k$ labels match the argument, one can show that
\begin{align}
&\frac{\dd}{\dd t_0} 
  \GLarg{n_1 &n_2 &\ldots &n_{k} &n_{k+1} &\ldots &n_r}{t_0 &t_0 &\ldots &t_0 &a_{k+1} &\ldots  &a_r}{t_0}  \label{tzero4}\\
&= \Bigg(  \prod_{j=1}^{k-1} \int \limits^{t_{j-1}}_0 \dd t_j \, f^{(n_j)}(t_j-t_0) \Bigg) 
   \int^{t_{k-1}}_0 \dd t \, f^{(n_k)}(t-t_0) f^{(n_{k+1})}(t-a_{k+1}) 
   \GLarg{n_{k+2} &\ldots &n_r}{a_{k+2} &\ldots  &a_r}{t}\,.\notag
\end{align}
For a terminal sequence of $a_j=t_0$, we find
\begin{align}
&\frac{\dd}{\dd t_0} \GLarg{n_1 &\ldots &n_{\ell-1} &n_{\ell} &\ldots &n_r}{a_1 &\ldots &a_{\ell-1} &t_0 &\ldots &t_0}{t_0} =
f^{(n_1)}(t_0-a_1)  \GLarg{n_2 &\ldots &n_{\ell-1} &n_{\ell} &\ldots &n_r}{a_2 &\ldots &a_{\ell-1} &t_0 &\ldots &t_0}{t_0}   \nnl
%%%
&- \Bigg(  \prod_{j=1}^{\ell-2} \int \limits^{t_{j-1}}_0 \dd t_j \, f^{(n_j)}(t_j-a_j) \Bigg) \int^{t_{\ell-2}}_0 \dd t \, f^{(n_{\ell-1})}(t-a_{\ell-1}) f^{(n_{\ell})}(t-t_{0}) \GLarg{n_{\ell+1} &\ldots &n_r}{t_0 &\ldots &t_0}{t}
\nnl
&+ f^{(n_r)}(-t_0) \GLarg{n_1 &\ldots &n_{\ell-1} &n_{\ell} &\ldots &n_{r-1}}{a_1 &\ldots &a_{\ell-1} &t_0 &\ldots &t_0}{t_0}\,. 
\label{tzero5}
\end{align}
Finally, an intermediate sequence of $a_{j}=t_0$ ranging from $j=p$ to $j=q$
with $p\neq 1$ and $q \neq r$ can be addressed via
\begin{align}
&\frac{\dd}{\dd t_0} \GLarg{n_1 &\ldots &n_{p-1} &n_{p} &\ldots &n_q &n_{q+1} &\ldots &n_r}{a_1 &\ldots &a_{p-1} &t_0 &\ldots &t_0 &a_{q+1} &\ldots &a_r}{t_0} =
f^{(n_1)}(t_0-a_1)
\GLarg{n_2 &\ldots &n_{p-1} &n_{p} &\ldots &n_q &n_{q+1} &\ldots &n_r}{a_2 &\ldots &a_{p-1} &t_0 &\ldots &t_0 &a_{q+1} &\ldots &a_r}{t_0} \nnl
%%%%%
&- \Bigg(  \prod_{j=1}^{p-2} \int \limits^{t_{j-1}}_0 \dd t_j \, f^{(n_j)}(t_j-a_j) \Bigg)  \notag \\
&\ \ \ \ \ \ \ \ \ \ \ \ \ \ \ \ \ \ \ \ \ \times
   \int^{t_{p-2}}_0 \dd t \, f^{(n_{p-1})}(t-a_{p-1}) f^{(n_{p})}(t-t_0)
   \GLarg{n_{p+1} &\ldots &n_q &n_{q+1} &\ldots &n_r}{t_0 &\ldots &t_0 &a_{q+1} &\ldots &a_r}{t} \nnl
%%%%%
&+ \Bigg(  \prod_{j=1}^{p-1} \int \limits^{t_{j-1}}_0 \dd t_j \, f^{(n_j)}(t_j-a_j) \Bigg)
   \Bigg(  \prod_{j=p}^{q-1} \int \limits^{t_{j-1}}_0 \dd t_j \, f^{(n_j)}(t_j-t_0) \Bigg) \label{tzero6}
\\
&\ \ \ \ \ \ \ \ \ \ \ \ \ \ \ \ \ \ \ \ \ \times \int^{t_{q-1}}_0 \dd t \, f^{(n_{q})}(t-t_{0}) f^{(n_{q+1})}(t-a_{q+1})   \GLarg{n_{q+2} &\ldots &n_r}{a_{q+2} &\ldots &a_r}{t} \ .
\notag
\end{align}
%%%%%%%%%%%%%%%%%%%%%%%%%%%%%%%%%%%%%%%%%%
Cases with multiple disconnected sequences of $a_j=t_0$ can be treated along
similar lines.

\subsection{Recursive removal of the argument from the labels}
\label{ellpolyrem}

On the basis of eqns.~\eqref{tzero4} to
\eqref{tzero6}, we can generalize the recursions eqns.~\eqref{extra21} to
\eqref{extra23} to situations where several successive instances
of the argument occur among the labels. If the first $k$ labels match the
argument, one can show that
\begin{align}
& \GLarg{n_1 &n_2 &\ldots &n_{k} &n_{k+1} &\ldots &n_r}{z &z &\ldots &z &a_{k+1} &\ldots  &a_r}{z} = \lim_{z\rightarrow 0} G(z,\ldots,z,a_{k+1},\ldots,a_r;z) \prod_{j=1}^r \delta_{n_j,1} \notag
\\
&-(-1)^{n_k}\int^z_0 \dd t \, f^{(n_k+n_{k+1})}(t-a_{k+1}) \GLarg{n_1 &\ldots &n_{k-1} &0 &n_{k+2} &\ldots &n_r}{t &\ldots &t &0 &a_{k+2} &\ldots &a_r}{t}  \label{tzero81} \\
&+ \sum_{j=0}^{n_{k+1}} {n_k-1+j \choose j} \int^z_0 \dd t \, f^{(n_{k+1}-j)}(t-a_{k+1}) \GLarg{n_1 &\ldots &n_{k-1} &n_k+j &n_{k+2} &\ldots &n_r}{t &\ldots &t &t &a_{k+2} &\ldots &a_r}{t} \notag \\
&+ \sum_{j=0}^{n_k} {n_{k+1}-1+j \choose j} (-1)^{n_k+j} \int^z_0 \dd t \, f^{(n_k-j)}(t-a_{k+1}) \GLarg{n_1&\ldots &n_{k-1} &n_{k+1}+j &n_{k+2} &\ldots &n_r }{t &\ldots &t &a_{k+1} &a_{k+2} &\ldots &a_r}{t}
\,.\notag
\end{align}
For a terminal sequence of $a_j=z$, we find
\begin{align}
& \GLarg{n_1 &\ldots &n_{\ell-1} &n_{\ell} &\ldots &n_r}{a_1 &\ldots &a_{\ell-1} &z &\ldots &z}{z} =  \lim_{z\rightarrow 0} G(a_{1},\ldots,a_{\ell-1},z,\ldots,z;z) \prod_{j=1}^r \delta_{n_j,1} \nnl
%%%
&\ \ \ \ + \int^z_0 \dd t \, f^{(n_1)}(t-a_1) \, \GLarg{n_2 &\ldots &n_{\ell-1} &n_{\ell} &\ldots &n_r}{a_2 &\ldots &a_{\ell-1} &t &\ldots &t}{t}
\nnl
&\ \ \ \ +(-1)^{n_\ell} \int^z_0 \dd t \, f^{(n_\ell+ n_{\ell-1})}(t-a_{\ell-1}) \, \GLarg{n_1 &\ldots &n_{\ell-2} &0 &n_{\ell+1} &\ldots &n_r}{a_1 &\ldots &a_{\ell-2} &0 &t &\ldots &t}{t} \label{tzero82} \\
&\ \ \ \ - \sum_{j=0}^{n_{\ell-1}} {n_\ell-1+j \choose j} \int^z_0 \dd t \, f^{(n_{\ell-1}-j)}(t-a_{\ell-1}) \GLarg{n_1 &\ldots &n_{\ell-2} &n_\ell+j &n_{\ell+1} &\ldots &n_r}{a_1 &\ldots &a_{\ell-2} &t &t &\ldots &t}{t} \notag \\
&\ \ \ \ -\sum_{j=0}^{n_{\ell}} {n_{\ell-1}-1+j \choose j} (-1)^{n_\ell+j} \int^z_0 \dd t \, f^{(n_\ell-j)}(t - a_{\ell-1}) \GLarg{n_1 &\ldots &n_{\ell-2} &n_{\ell-1}+j &n_{\ell+1} &\ldots &n_r}{a_1 &\ldots &a_{\ell-2} &a_{\ell-1} &t &\ldots &t}{t} \notag \\
&\ \ \ \ + (-1)^{n_r} \int^z_0 \dd t \,f^{(n_r)}(t) \GLarg{n_1 &\ldots &n_{\ell-1} &n_{\ell} &\ldots &n_{r-1}}{a_1 &\ldots &a_{\ell-1} &t &\ldots &t}{t}
\,. \notag
\end{align}
Finally, an intermediate sequence of $a_{j}=z$ ranging from $j=p$ to $j=q$
with $p\neq 1$ and $q \neq r$ can be addressed via
\begin{align}
& \GLarg{n_1 &\ldots &n_{p-1} &n_{p} &\ldots &n_q &n_{q+1} &\ldots &n_r}{a_1 &\ldots &a_{p-1} &z &\ldots &z &a_{q+1} &\ldots &a_r}{z} =
\lim_{z\rightarrow 0} G(a_{1},\ldots,a_{p-1},z,\ldots,z,a_{q+1},\ldots, a_r;z)
\prod_{j=1}^r \delta_{n_j,1} \nnl
%%%%%
&+ \int^z_0 \dd t \, f^{(n_1)}(t-a_1) \GLarg{n_2 &\ldots &n_{p-1} &n_p &\ldots &n_q &n_{q+1} &\ldots &n_r}{a_2 &\ldots &a_{p-1} &t &\ldots &t &a_{q+1} &\ldots &a_r}{t}
\nnl
%%%%%
&+ (-1)^{n_p} \int^z_0 \dd t \, f^{(n_p+n_{p-1})}(t-a_{p-1}) \GLarg{n_1 &\ldots &n_{p-2} &0 &n_{p+1} &\ldots &n_q &n_{q+1} &\ldots &n_r}{a_1 &\ldots &a_{p-2} &0 &t &\ldots &t &a_{q+1} &\ldots &a_r}{t}
\label{tzero83}
\\
&- \sum_{j=0}^{n_{p-1}} { n_p - 1 + j \choose j} \int^z_0 \dd t \, f^{(n_{p-1}-j)}(t-a_{p-1}) \GLarg{n_1 &\ldots &n_{p-2} &n_p+j &n_{p+1} &\ldots &n_q &n_{q+1} &\ldots &n_r}{a_1 &\ldots &a_{p-2} &t &t &\ldots &t &a_{q+1} &\ldots &a_r}{t} \notag \\
&-\sum_{j=0}^{n_p} {n_{p-1}\!-\!1\!+\!j \choose j} (-1)^{n_p+j} \! \int^z_0 \! \dd t \, f^{(n_p-j)}(t-a_{p-1})\GLarg{n_1 &\!\ldots \! &n_{p-2} &n_{p-1}+j &n_{p+1} &\!\ldots \!&n_q &n_{q+1} &\!\ldots \! &n_r}{a_1 &\!\ldots \!&a_{p-2} &a_{p-1} &t &\!\ldots \!&t &a_{q+1} &\!\ldots \!&a_r}{t} \notag \\
&- (-1)^{n_q} \int^z_0 \dd t \, f^{(n_q+n_{q+1})}(t-a_{q+1}) \GLarg{n_1 &\ldots &n_{p-1} &n_p &\ldots &n_{q-1} &0 &n_{q+2} &\ldots &n_r }{a_1 &\ldots &a_{p-1} &t &\ldots &t &0 &a_{q+2} &\ldots &a_r}{t} \notag \\
&+\sum_{j=0}^{n_{q+1}} {n_q-1+j \choose j} \int^z_0 \dd t \,f^{(n_{q+1}-j)}(t-a_{q+1}) \, \GLarg{n_1 &\ldots &n_{p-1} &n_p &\ldots &n_{q-1} &n_q+j &n_{q+2} &\ldots &n_r}{a_1 &\ldots &a_{p-1}&t &\ldots &t &t &a_{q+2} &\ldots &a_r}{t} \notag \\
&+ \sum_{j=0}^{n_q} {n_{q+1} \! -\!1\!+\! j \choose j} (-1)^{n_q+j}\! \int^z_0 \! \dd t \, f^{(n_q-j)}(t-a_{q+1})  \GLarg{n_1 &\!\ldots\! &n_{p-1} &n_p &\!\ldots \! &n_{q-1} &n_{q+1}+j &n_{q+2} &\ldots &n_r}{a_1 &\!\ldots\! &a_{p-1} &t &\!\ldots \!&t &a_{q+1} &a_{q+2} &\!\ldots\! &a_r}{t}
 \ .
\notag
\end{align}
These relations reproduce eqns.~\eqref{extra21} to \eqref{extra23} for $k=1$,
$p=q$ and $\ell=r$, respectively.

\subsection{Eliminating labels \texorpdfstring{$a_j=z$}{ajz} at length three}
\label{lthree}

The generalization of \eqn{l2.3a} to length three is governed by
\begin{align}
&\GLargz{n_1 &n_2 &n_3}{z&0&0} =   -\zeta_3  \delta^1_{n_1}\delta^1_{n_2}\delta^1_{n_3}+ 
\zeta_2 \sum_{ j=0}^{n_2} \delta_{n_3}^1 \delta^1_{n_1+j} {n_1 - 1 + j\choose j}  \GL(n_2 - j;z)   \notag \\
&- (-1)^{n_1} \GL(n_1 + n_2, 0, n_3;z)+\sum_{j=0}^{n_1}  (-1)^{n_1 + j} {n_2 - 1 + j\choose j} \GL(n_1 - j, n_2 + j, n_3;z)  \notag \\
& - 
 \sum_{j=0}^{n_2} (-1)^{n_1 + j} {n_1 - 1 + j\choose j} \GL(n_2 - j, n_1 + n_3 + j, 
    0;z)  \label{sumexp2} \\
    & + \sum_{j=0}^{n_2}  {n_1 - 1 + j\choose   j} \sum_{k=0}^{n_3} (-1)^{n_1 + j + k}{n_1 + j - 1 + k \choose k} \GL(n_2 - j, 
      n_3 - k, n_1 + j + k ;z) \notag \\
      & + \sum_{j=0}^{n_2} {n_1 - 1 + j\choose    j} \sum_{k=0}^{n_1+j} (-1)^{n_1 + j + k} {n_3 - 1 + k\choose k} \GL(n_2 - j, 
      n_1 + j - k, n_3 + k;z)  \ .
    \notag
\end{align}
The reflection identity (\ref{eqn:reflection}) allows to infer $\GLargz{n_1 &n_2 &n_3}{z&z&0}= (-1)^{n_1+n_2+n_3}\GLargz{n_3 &n_2 &n_1}{z&0&0}$, and permutations in the labels are covered by shuffle relations.

\section{Trigonometric integrals}
\label{app:trig}

This appendix gathers trigonometric integrals relevant for the evaluation of
eMZVs. The result in \eqn{explq1} for $\omega(0,1,0,0)$ relies on 
\begin{align}
\int^1_0 \dd z_4 \int^{z_4}_0 \dd z_3\int^{z_3}_0 \dd z_2 \, \sin(2\pi n z_2) \, z_2 &= \frac{ 3}{8\pi^3 n^3} \label{explq4} \\
\int^1_0 \dd z_4 \int^{z_4}_0 \dd z_3\int^{z_3}_0 \dd z_2 \, \cot(\pi  z_2) \, z_2 &= \frac{3\zeta_3}{4\pi^3} \ ,
\label{explq5}
\end{align}
and the eMZVs relevant at order $s_{ij}^2$ as given by \eqn{explq2} and
\eqn{explq3} are based on
\begin{align}
\int^1_0 \dd z_5 \int^{z_5}_0 \dd z_4\int^{z_4}_0 \dd z_3 \int^{z_3}_0 \dd z_2 \int^{z_2}_0 \dd z_1 \, \cos(2\pi n z_1)
&= \frac{ 1}{24\pi^2 n^2} - \frac{ 1}{16\pi^4 n^4} \label{explq6}\\
\int^1_0 \dd z_5 \int^{z_5}_0 \dd z_4\int^{z_4}_0 \dd z_3 \int^{z_3}_0 \dd z_2 \, \cos(2\pi n z_2) \, z_2 &= - \frac{ 1}{24\pi^2 n^2} + \frac{ 1}{4\pi^4 n^4} \ .
\label{explq7}
\end{align}

\section{Cycle index of the symmetric group and the \texorpdfstring{$\boldsymbol{f^{(n)}}$}{fn} functions}
\label{app:cyc}

This appendix highlights the connection between the explicit expansion of the
doubly-periodic functions $f^{(n)}$ in \eqref{fns} with the cycle index of the
symmetric group $S_n$. For general references, see \cite{Cameron,Riordan}.

\paragraph{Cycle structures.} Every permutation $g\in S_n$ of $X=\{1,
\ldots,n\}$ can be written as the product of disjoint cycles with lengths $a_1,
\ldots,a_n$ such that $n=\sum_{i=1}^n a_i$.  This integer partition of $n$ is
represented by $\lambda = 1^{a_1}2^{a_2} \ldots n^{a_n}$ and is called the
cycle structure of the permutation. Therefore the total number of cycle
structures for the permutations in $S_n$ is given by the integer partition
$P(n)=1,2,3,5,7, \ldots$. Note that the number of terms in each $f^{(n)}$ is
also $P(n)$.  Furthermore, if $\lambda = 1^{a_1}2^{a_2} \ldots n^{a_n}$ is a
partition of $n$ (denoted by $\lambda \vdash n$), the number of permutations
with cycle structure $\lambda$ is \cite{Cameron}
\begin{equation}
\label{cycleN}
{n!\over \prod_{i=1}^n i^{a_i}a_i!}\,.
\end{equation}
Note that the coefficients of the monomials $\cE_1^{a_1} \ldots\cE_n^{a_n}$ in
$f^{(n)}$ given by \eqn{alt21} are reproduced by the formula \eqref{cycleN}
with the corresponding cycle structure.  This observation can be made more
precise with the definition of the \textit{cycle index} of the symmetric group
$S_n$ \cite{Cameron},
\begin{equation}
\label{cycInd}
Z(S_n;s_1, \ldots,s_n) = {1\over n!}\sum_{g\in S_n} z(g; s_1, \ldots,s_n) \ ,
\end{equation}
where $z(g; s_1, \ldots,s_n) = s_1^{a_1}s_2^{a_2} \ldots s_n^{a_n}$ and $a_i$
counts the number of cycles of length $i$ in the permutation $g$. One can see
from the first few examples\footnote{In addition, it is convenient to define
$Z(S_0) = 1$.},
\begin{align}
\label{ZSn}
Z(S_1, s_1) &= s_1 \cr
Z(S_2, s_1, \ldots,s_2) &= {1\over 2!}\big( s_1^2 + s_2\big)\notag\cr
Z(S_3, s_1, \ldots,s_3) &= {1\over 3!}\big( s_1^3 + 3s_1s_2 + 2 s_3\big)\notag\cr
Z(S_4, s_1, \ldots,s_4) &= {1\over 4!}\big( s_1^4 + 6s_1^2s_2 + 8 s_1 s_3 + 3s_2^2 + 6 s_4\big)\notag
\end{align}
that the cycle index of $S_n$ captures the expansions in \eqref{fns}. More
precisely, theorem 1.3.3 of \cite{StanleyBook} can be written as
\begin{equation}
\label{PropCyc}
\sum_{n=0}^\infty \alpha^n Z(S_n; \cE_1, \ldots,\cE_n) = \exp\Big(\sum_{j=1}^\infty  {\cE_j\over j}
\alpha^j\Big)\,,
\end{equation}
and comparing \eqref{calee2} with \eqref{PropCyc} leads to,
\begin{align}
\label{fnasCyc}
f^{(n)} &= Z(S_n; \cE_1, \ldots,\cE_n)\cr
& = \sum_{\lambda \vdash n} \prod_{i=1}^n {\cE_i^{a_i}\over i^{a_i}a_i!},\quad \lambda = 1^{a_1}2^{a_2}
\ldots n^{a_n}\,.
\end{align}
Furthermore, one can also show that \cite{Riordan},
\begin{equation}
\label{delf}
{\partial f^{(n)}(z,\tau)\over \partial \cE_p} = {1\over p}f^{(n-p)}(z,\tau).
\end{equation}
Note, in particular, that \eqref{delf} yields an alternative proof of
\eqref{alt12},
\begin{equation}
\label{altern}
{\partial f^{(n)}(z,\tau)\over \partial {\bar z}} = {\partial f^{(n)}(z,\tau)\over \partial
\cE_1}{\partial \cE_1 \over \partial {\bar z}} = - {\pi\over \text{Im}(\tau)}f^{(n-1)}(z,\tau).
\end{equation}

\paragraph{Symmetric polynomials.} The cycle index of the symmetric group $S_n$
also provides a recipe for expressing the \textit{complete symmetric function}
$h_j$ in terms of the \textit{power sum function} $p_j$, i.e., $h_n=Z(S_n;
p_1,p_2, \ldots,p_n)$ \cite{Cameron}. Therefore the functional form of $h_n$
matches that of $f^{(n)}$ and one can exploit the well-known relation $nh_n =
\sum_{i=1}^n p_i h_{n-i}$ from the theory of symmetric functions to obtain the
corresponding recursion formula \eqn{alt22} for $f^{(n)}$.

\bibliographystyle{nb}
\bibliography{\jobname}

\end{document}